\documentclass[a4paper,11pt]{article}
\pdfoutput=1  
\usepackage{jcappub}  
\usepackage[T1]{fontenc}  
\usepackage{amssymb,amsfonts,slashed,amsthm,amsmath,graphicx}
\usepackage[all]{xy}
\usepackage{dcolumn}
\usepackage{bm}
\usepackage{mathrsfs}
\usepackage{ucs}
\usepackage[utf8x]{inputenc}
\usepackage{hyperref}
\usepackage{hepunits}
\usepackage{slashed}

\newcommand{\be}{\begin{equation}}
\newcommand{\ee}{\end{equation}}
\newcommand{\bea}{\begin{eqnarray}} 
\newcommand{\eea}{\end{eqnarray}}
\newcommand{\ft}[2]{{\textstyle\frac{#1}{#2}}}

\def\pd{\ensuremath{\partial}}

\def\rme{{\mathrm e}}
\def\rmi{{\mathrm i}}

\newsavebox{\uuunit}
\sbox{\uuunit}
    {\setlength{\unitlength}{0.825em}
     \begin{picture}(0.6,0.7)
        \thinlines
        \put(0,0){\line(1,0){0.5}}
        \put(0.15,0){\line(0,1){0.7}}
        \put(0.35,0){\line(0,1){0.8}}
       \multiput(0.3,0.8)(-0.04,-0.02){12}{\rule{0.5pt}{0.5pt}}
     \end {picture}}

\newcommand{\U}{\mathop{\rm {}U}}

   \def\cD{{\cal D}}

  \def\cN{{\cal N}}

\def\atH0{|_{H_{0}}}
\def\AtH0{\bigg|_{H_{0}}}

\csname @addtoreset\endcsname{equation}{section}

\title{\huge{Bubbles of nothing and
supersymmetric compactifications}}
\vspace{-3cm}
\author[1,2]{\large{Jose J. Blanco-Pillado,}}
\author[3,4]{Benjamin Shlaer,}
\author[2,5]{Kepa Sousa}
\author[2]{and Jon Urrestilla}
\affiliation[1]{IKERBASQUE,  Basque Foundation for Science, 48011, Bilbao, Spain}
\affiliation[2]{Department of Theoretical Physics, University of the Basque Country UPV/EHU,\\
48080 Bilbao, Spain}
\affiliation[3]{Department of Physics, University of Auckland, Private Bag 92019, Auckland, New Zealand}
\affiliation[4]{Institute of Cosmology, Department of Physics and Astronomy, Tufts University, Medford,  MA 02155, USA}
\affiliation[5]{Instituto de Fisica Teorica UAM-CSIC
Universidad Autonoma de Madrid, Cantoblanco, 28049 Madrid, Spain}

\emailAdd{josejuan.blanco@ehu.eus}
\emailAdd{kepa.sousa@csic.es}
\emailAdd{shlaer@cosmos.phy.tufts.edu}
\emailAdd{jon.urrestilla@ehu.eus}

\abstract{We investigate the non-perturbative stability of supersymmetric compactifications with 
respect to decay via a bubble of nothing.  We show examples where this kind of instability 
is not prohibited by the spin structure, i.e., periodicity of fermions about the extra dimension. 
However, such ``topologically unobstructed'' cases do exhibit an extra-dimensional analog of the well-known Coleman-De Luccia 
suppression mechanism, which prohibits the decay of supersymmetric vacua.
We demonstrate this explicitly in a four dimensional Abelian-Higgs toy model coupled to supergravity. The 
compactification of this model to $M_3 \times S_1$ presents the possibility of vacua with different 
windings for the scalar field.  Away from the supersymmetric limit, these states decay by the formation
of a bubble of nothing, dressed with an Abelian-Higgs vortex.   
We show how, as one approaches the supersymmetric limit, the
circumference of the topologically unobstructed bubble becomes infinite, thereby preventing the realization of this decay. 
This demonstrates the dynamical origin of the decay suppression, as opposed to the more familiar argument based on 
the spin structure.  We conjecture that this is a generic mechanism that enforces stability of any topologically unobstructed supersymmetric compactification.
}

\begin{document}
\maketitle
\flushbottom

\section{Introduction}
\label{sec:intro}

Higher dimensional theories have been extensively studied in the last few decades
as possible extensions of the Standard Model.  Incorporating these extra dimensions
into the fundamental theory adds new degrees of freedom to the low energy
dynamics.  One must first demonstrate the existence of a
perturbatively stable vacuum in such a theory before one can 
infer any new physics. This is normally achieved by a compactification
mechanism that generates a potential for these new 
degrees of freedom. Our vacuum might be a minimum in such a potential,
and therefore perturbatively stable.  Finding stable vacua compatible with current observations is
one of the major challenges facing all fundamental higher dimensional theories, such as String Theory. 

It is far from obvious that there should exist a unique vacuum in the effective
4D description of these higher dimensional theories, and often one encounters multiple minima within the same
compactification potential. This opens up the possibility of a new type of instability due to 
the presence of large quantum mechanical fluctuations about the
original vacuum. Such tunneling processes were first discussed in a series of 
papers \cite{Coleman:1977py,Callan:1977pt} in the context of a 4D field theory. The 
results of these papers
are that a non-perturbative instability of false vacua occurs through the nucleation
of a bubble, whose interior consists of the lower energy vacuum.  
The bubble wall is a field configuration interpolating 
between the higher energy initial vacuum (parent vacuum) and the final
one (daughter vacuum).  In many cases, one can reduce this problem to the 
simpler version within the so-called thin wall approximation, where one can
assume that the bubble is spherical and made of a thin solitonic domain wall (of fixed tension) characterized by just its radius.  
Once the bubble is formed by the quantum mechanical tunneling process, 
the pressure difference across the wall accelerates the
bubble, making it grow and gain kinetic energy by converting arbitrarily large regions 
of spacetime into the new lower energy density vacuum.  Clearly it is important to
estimate the rate of this instability if our universe is to be
described by any of the theories susceptible to vacuum decay.  One
can calculate the rate of this decay by the use of instanton methods
and the computation of the action of the appropriate Euclidean classical solutions. 
This was done in \cite{Callan:1977pt} and further developed in the 
thin wall approximation for the case of a scalar field potential coupled
to gravity in \cite{Coleman:1980aw}.  After coupling the theory to gravity, one discovers that
there are now cases where the tunneling probability is completely suppressed.
Certain Minkowski and Anti-deSitter false
vacua are exactly stable even at the non-perturbative level. 
The reason for this is that all saddle points of their Euclidean action correspond to
bubbles of infinite circumference, which thus have infinite action.  The nucleation rate is 
exponentially suppressed by the Euclidean action, and therefore vanishes.  This enforces the stability of the 
parent vacuum, making such states interesting starting 
points for searches for realistic vacua in theories beyond the Standard Model.

Another common ingredient in many extensions of the Standard Model is supersymmetry. 
In particular, many higher dimensional theories also incorporate 
supersymmetry (such as in string theory).  It is therefore natural to consider compactifications 
that preserve some supersymmetry. This is not only interesting from the point of view of
phenomenology, but also makes the question of stability a much simpler one
to study, at least perturbatively \cite{BlancoPillado:2005fn}.

 At the non-perturbative level the
question of stability becomes much more interesting. This was studied in
the context of a simple $d=4$,  $\cN=1$ supergravity model in \cite{Cvetic:1992st}. The results of that paper indicate
that indeed the supersymmetric vacua are stabilized by the previously mentioned Coleman-De Luccia suppression
mechanism. The conditions enabling suppression are enforced in that
case by the form of the potential: in a theory with distinct supersymmetric vacua,
there is a solution containing a domain wall interpolating between them.
This domain wall is static and has infinite area.
One way to understand Coleman-De Luccia suppression is to think about the required tension
of the bubble wall that one would need in a thin wall description of the decay.  
For sufficiently low tension, the Euclidean action of the thin wall bubble always has a saddle point for a certain critical value of
bubble radius, describing the size of the bubble at nucleation.  As the tension is increased, the critical radius grows,
and at a certain finite value of tension, the critical radius diverges.
The static domain wall described above is described precisely by
this saddle point, and the infinite area is due to the domain wall having the tension corresponding to a divergent critical radius.

From this analysis one concludes that there is a limiting tension above which the
decay process cannot happen.  Interestingly, supersymmetry imposes a lower
bound on the tension of the wall interpolating between supersymmetric vacua,
so it is clear that the bubble decay process cannot occur if this lower bound on tension is at or above the tension corresponding to an infinite critical radius. 
Furthermore, one can show that the limiting case where these two values of tension coincide preserves
part of the original supersymmetry.

 These arguments seem to suggest that models of compactification which can be
described by a supersymmetric theory would be stable if their compactification
preserves part of the supersymmetry. However, there may be new instabilities 
of a higher dimensional theory that are not described in the context of a
4D low energy effective theory and therefore one would have to take all the previous 
considerations with a little bit of caution. An example of such an instability was demonstrated
some time ago by Witten \cite{Witten:1981gj}.  In the simplest example of a 5D Kaluza-Klein
compactification
to $\mathbb{M}_4\times S^1$, 
he was able to explicitly construct an instanton for the decay
by the formation of a {\it bubble of nothing}. This instanton describes 
the formation of a bubble where the extra dimensions pinch off, disappearing, signaling the end
of spacetime in this region, hence the name bubble of nothing. Viewed from a 4D perspective, these
solutions would be singular\footnote{See for example the discussion in \cite{Dine:2004uw}.}, so it is hard to argue their existence or validity on the basis
of a pure 4D theory. Nevertheless, these potential new instabilities exist and 
one would wonder if this could lead to the decay of some supersymmetric compactifications.
 The original paper by Witten  \cite{Witten:1981gj} already contains clues regarding this subject, which lead to the conclusion
  that it would be impossible for a supersymmetric compactification with a circle extra dimension to decay
in this way. The argument is quite simple. The 5D Kaluza-Klein theory allows for
a supersymmetric extension including fermionic degrees of freedom, but its compactification would
only preserve some supersymmetry if these fermionic modes are periodic around the 
extra dimension. On the other hand, the instanton solution that allows the decay
and disappearance of the extra-dimension (into nothing) forces the situation with
anti-periodic fermions, so it is clear that one would be in a different sector of the theory
if one starts with a supersymmetric compactification and this decay would not be 
possible\footnote{For a discussion of the pre-factor of the decay probability in this context see \cite{Rubakov:2012tj}.}.

In this paper we want to investigate how generic this argument is. In particular,
the justification for stability of supersymmetric compactifications in Witten's argument 
is entirely based on the spin structure of the theory and the 
instanton solution and it seems hard to generalize it to other 
internal spaces. Furthermore, the reasoning for the tunneling suppression is also quite different in nature
from the one described earlier in the $\cN=1$ supergravity scalar theory that
relies on a dynamical mechanism first identified
in field theory by Coleman and De Luccia \cite{Coleman:1980aw}.  

The main idea of this paper is to look for the simplest example of a supersymmetric
theory where the  argument for stability based on the spin structure cannot be used and investigate in this
case the possible existence of a bubble of nothing instability. In the following,
we will show that there are indeed some supersymmetric compactifications that allow
for the same spin structure as the bubble of nothing geometry, therefore circumventing
Witten's argument on stability. However, we will show that in these cases, the stability
of the compactification is preserved by the Coleman-De Luccia suppression mechanism,
where the nucleated bubble would need to be infinitely large and the decay would therefore be completely
suppressed. Hence our conclusions are that, in fact, supersymmetric compactifications
remain stable but that the reason for this in some cases may be different from what was originally 
envisioned in \cite{Witten:1981gj}.\\

The plan of the paper is the following. We describe in section \ref{sec:model} the simple model 
of $d=4$, $\cN=1$ supergravity that we will consider.  We show in section \ref{sec:compactification} how one 
can compactify this theory on a circle down to three dimensions, analogous to 
the usual Kaluza-Klein model. We also show in this section the conditions required to
obtain a supersymmetric compactification of this model and study the spin structure
of those vacuum solutions. We discuss in section \ref{sec:bubblewithstring} the kind of instanton solutions
one would need in order to describe the decay of these compactified vacuum states
and their relation to the original bubbles of nothing in the thin wall approximation. 
In section \ref{sec:AHBON}, we present our numerical approach to
study these instanton solutions in the supergravity model presented earlier.
In section \ref{sec:thinwall} we compare analytic and numerical solutions 
within the thin wall regime and carefully investigate their limit as the initial state 
becomes supersymmetric. In section \ref{sec:thickwall} we present generic numerical solutions. 
Finally we conclude with some remarks in section \ref{sec:conc}.

\section{The model}
\label{sec:model}
 
As we mentioned in the introduction, we would like to study a model whose compactification on a circle
allows for anti-periodic fermionic boundary conditions, while still preserving part of its supersymmetry. We 
will show in this paper that we can find a compactification with these characteristics within the
 Abelian-Higgs model coupled to  $\cN=1$ supergravity in $3+1$ dimensions.\footnote
 {We make this choice in order to 
simplify the model and to use the well known 4D supersymmetric notation, but we do expect the results of this paper to apply to 
more general models. It would be interesting to look for generalizations of this
idea to a truly higher dimensional model or models with other matter content.} 
This model has been considered in the literature
mainly in the context of cosmic string solutions \cite{Edelstein:1995ba,Dvali:2003zh} and we will see later on that
these solutions also play a role in our current discussion.\footnote
{In this paper, we will follow the conventions in \cite{Freedman:2012zz}. In particular we use the Minkowski metric with 
signature $(-,+,+,+)$, and  we work with the units $c=\hbar=1$,  so that  the reduced Planck mass reads 
$\kappa^2 \equiv M_P^{-2}= 8 \pi G$.}

 The model  describes the dynamics of a complex scalar field $\phi$  with K\"ahler potential 
 $K(\phi ,\bar \phi) =\bar \phi \phi$ minimally coupled to a $\mathrm{U}(1)$ gauge field $A_\mu$. The 
 superpotential is taken to be $W(\phi)=0$, and the gauge kinetic function is $f(\phi)=1$. 
 With these choices, the bosonic part of the action is the well known  Einstein-Abelian-Higgs model
 \begin{eqnarray}
\label{action}
 S_{\textrm{bos}} &=&  \int d^4x \sqrt{- g} \Big[  \frac{1}{2\kappa^2} \, R -  D_{\mu} \bar \phi  \,  D^{\mu} \phi -\frac{1}{4} 
F_{\mu\nu} F^{\mu\nu} -\frac{e^2}{2}\left(\eta^2 -  \phi \bar\phi \right)^2
\label{phenomL}
\Big], 
\end{eqnarray}
 where the gauge covariant derivative is defined by $D_{\mu} \phi=(\partial_{\mu} - i e A_{\mu}) \phi$, and   
 $F_{\mu\nu}=\partial_{\mu} A_{\mu} - \partial_{\nu} A_{\mu}$ is the $\mathrm{U}(1)$ field strength.  We also introduce for future reference
the dimensionless parameter $\gamma  \equiv \kappa^2 \eta^2$, which controls, as we will see later on, the gravitational
effects of the typical energy scale of this theory.

The full  supergravity model also involves  the  gravitino field, $\psi_\mu$, and the fermionic partners of the chiral 
 and gauge fields, $\chi$ and $\lambda$, respectively.\footnote
 	{The gravitino is usually written as a Majorana spinor $\psi_\mu $, but sometimes it is convenient to split it into its complex chiral parts,
	$\psi _{\mu L}=\ft12(1+\gamma _5)\psi _\mu$,  and $\psi _{\mu R}=\ft12(1-\gamma _5)\psi _\mu$.  
	The same notation applies to the gauginos $\lambda$ and the chiralinos $\chi$.}
It  is invariant under the  \emph{local} $\U(1)$ gauge transformations 
\bea
&&\delta_g{\phi} = \rmi e \phi \, \alpha\,, \nonumber\\
 &&  \delta_g \chi_L = \rmi e \left(1+ \ft{\eta^2 \kappa^2}{2} \right) \chi_L \,  \alpha\,, \nonumber\\ 
&&\delta_g \psi _{\mu L} = - \rmi    e\ft{   \eta^2 \kappa^2}{2}   \psi_{\mu L} \, \alpha \,,\nonumber\\
&&\delta_g \lambda_L =-   \rmi e\ft{   \eta^2 \kappa^2}{2}  \lambda_L \, \alpha, 
\label{gauginoGtrans} 
\eea
where $\alpha$ is the gauge parameter. Note that the combination  $\xi\equiv e \eta^2$  appearing in the 
scalar potential also contributes to the charge of all fermions under the local $\U(1)$ symmetry. Such a 
combination can be identified as the  Fayet-Iliopoulos (FI) term of $\cN=1$ supergravity, and it is associated 
to  the gauging of the $R-$symmetry which rotates the supercharges. In order to simplify the notation, we will
take the FI term to be a free parameter in the main part of the paper and comment on its quantization 
in Appendix \ref{secondappendix}. The conclusions of the paper are not affected by this quantization.

In later sections we will discuss the spontaneous breaking of supersymmetry by bosonic backgrounds, 
so it will be useful to have the form of the supersymmetry transformations for this model. In purely  bosonic backgrounds  
only the supersymmetry transformations  of the fermions can be non-vanishing, which read
\bea
\delta \psi _{\mu L} &=& \cD_\mu \epsilon_L =  (\partial_\mu + \ft14 \omega_\mu^{ab} \gamma_{ab} + \ft{\rmi}{2} A_\mu^B ) \epsilon_L\,, \label{gravitinoSUSYtrans}\\
    \delta \chi_{L} &=& \ft{1}{\sqrt{2}} \gamma^\mu D_\mu \phi\,  \epsilon_R \,,  \label{chiralSUSYtrans}\\ 
\delta \lambda &=& \ft14 \gamma^{\mu\nu} F_{\phantom{a}\mu\nu} \epsilon +     \ft{\rmi}{2} e(\eta^2 -  \phi \bar \phi )\gamma_5\,   \epsilon,
\label{gauginoSUSYtrans} 
\eea
 up to terms cubic in the fermions.\footnote{See \cite{Freedman:2012zz} for further details.}
  Here $\epsilon$ is the parameter of the local supersymmetry transformations, 
 and the composite  $\U(1)$ connection $A_\mu^B$ is given by
\begin{equation}
  A_\mu ^B=\ft12 \rmi \kappa^2 \left[ \phi D_\mu \bar \phi -\bar \phi D_\mu \phi \right]
  + e \eta^2   \kappa^2 A_\mu \,.
  \label{AmuBinz}
\end{equation}

\section{Supersymmetric Kaluza-Klein compactification on $S^1$}
\label{sec:compactification}

\subsection{Generalized Kaluza-Klein compactification}

The model described by the action  \eqref{phenomL} admits a Kaluza-Klein type of vacuum, where one of the 
spatial dimensions is compactified on a circle, $\mathbb{M}_4 \to \mathbb{M}_3 \times S^1,$ and the matter fields
are in their vacuum, namely
\bea
\phi \bar \phi&=& \eta^2, \qquad D_\mu \phi =0, \qquad \psi_\mu = \chi = \lambda =0, \label{KKansatz}\\
 ds^2 &=& -dt^2 + dz^2 + dr^2 + R^2 d\theta^2. \label{KKansatzmetric}
\eea

Here the coordinates $t,z,r \in \mathbb{R}$ parametrize the 3-dimensional Minkowski part of the spacetime,
 and  $\theta \in [0,2 \pi)$ is the angular variable associated to the compact dimension whose physical circumference is
$2\pi R$. In order to define  the (generalized) Kaluza-Klein theory, it is necessary to specify the boundary conditions\footnote
{In general, the fields can be identified with sections on a non-trivial fibre bundle, 
and the choice of boundary conditions  specifies the topology of the bundle 
\cite{Isham:1977yc,Avis:1978qc,Isham:1978ec,Avis:1978tc,DeWitt:1979dd,Nakahara:2003nw}.
} 
of the fields 
around the compactified dimension \cite{Scherk:1978ta,Scherk:1979zr,Ortin:2015hya}. The fields need to be periodic only up to a global 
symmetry of the action, so we can formally write
\be
\Phi(\theta + 2 \pi) = \rme^{\rmi \hat Q \alpha}\,  \Phi (\theta),
\ee
 where we have denoted  all fields collectively by $\Phi$, and $\hat Q$ is the generator of a global symmetry.  Any Lorentz 
 invariant action is always invariant under the $\mathbb{Z}_2$ symmetry which flips the sign of all fermions. In addition, the 
 supergravity model  \eqref{phenomL} has a $\U(1)_c \times \U(1)_R$ global symmetry, where the first factor corresponds 
 to the $\U(1)$ symmetry associated with the chiral supermultiplet, and the  second one is the $R-$symmetry. Thus, on the spacetime 
 described by the line element \eqref{KKansatzmetric} we can impose boundary conditions of the form
\bea
\phi(\theta + 2 \pi)   \quad &=& \quad   \rme^{\rmi \alpha_c} \, \phi(\theta)\,,\nonumber\\
 \chi_L(\theta + 2 \pi) \quad &=&\quad \pm \, \rme^{ \rmi  \alpha_c}\;   \rme^{ \rmi   \alpha_R   }\,  \chi_L(\theta)\,,\nonumber\\ 
 \psi _{\mu L}(\theta + 2 \pi)  \quad &=&  \quad\pm \,  \rme^{- \rmi     \alpha_R} \, \psi_{\mu L}(\theta)\,,\nonumber  \\
  \lambda_L(\theta + 2 \pi)  \quad &=&\quad \pm \, \rme^{-   \rmi \alpha_R} \, \lambda_L(\theta)\,, 
\label{generalBC} 
\eea
where $\alpha_c\in[0,2\pi)$ and $\alpha_R\in[0,2 \pi)$ are the parameters of the global $\U(1)_c$ and $\U(1)_R$ respectively.  

As discussed in \cite{Witten:1981gj}, in a spacetime with the topology of a bubble of nothing,  the spinor fields are uniquely  
defined and only admit anti-periodic boundary conditions along the compactified dimension:\footnote{The construction of spinor structures on simply connected 
spacetimes and on the cylinder \eqref{trivialCylinder} is discussed in \cite{Isham:1978ec,DeWitt:1979dd,Ford:1979ds,Ford:1979pr,Ashtekar:2014ife}.}
\be
\chi_L(\theta + 2 \pi)  = -   \chi_L(\theta), \qquad 
 \psi _{\mu L}(\theta + 2 \pi) =-  \psi_{\mu L}(\theta),  \qquad  
 \lambda_L(\theta + 2 \pi)  =- \lambda_L(\theta).
\label{APbc} 
\ee

This implies that for the arbitrary boundary conditions of 
eqs.~\eqref{generalBC},  the Kaluza-Klein vacuum and the bubble of nothing belong to topologically 
distinct sectors.  This guarantees the generic stability of Kaluza-Klein vacua with respect to this decay channel \cite{Witten:1981gj}. 
In summary, only Kaluza-Klein vacua whose fermions satisfy anti-periodic boundary conditions may decay via the formation
of  bubbles of nothing.

\subsection{Pure vacuum solutions and periodic fermions}

For the Kaluza-Klein background to preserve the full supersymmetry of the model, the supersymmetry 
transformations  \eqref{gravitinoSUSYtrans}, \eqref{chiralSUSYtrans}  and \eqref{gauginoSUSYtrans}   must vanish for all values of the parameter $\epsilon$. In a background 
of the form  \eqref{KKansatz}, where the matter fields are on a pure vacuum configuration
\be
\phi = \eta, \qquad A_\mu =0, \qquad \psi_\mu = \chi = \lambda =0, \qquad ds^2 = -dt^2 + dz^2 + dr^2 + R^2 d\theta^2,
\label{trivialCylinder}
\ee
 only the gravitino transformation is non-trivial, which reduces to 
\be
\cD_\mu \epsilon_L =  \partial_\mu   \epsilon_L =0.
\ee

Then, for this background to preserve supersymmetry the theory must admit a  covariantly constant spinor. Such 
a solution must be globally well defined, meaning that it should be consistent with the boundary conditions \eqref{generalBC} that 
we have imposed for the Kaluza-Klein reduction. The solutions to the previous equation are just constant spinor 
parameters, $\epsilon_L (x^\mu) = \epsilon_{L}^0$, which are periodic, and thus 
the  background \eqref{trivialCylinder} can only be supersymmetric when the boundary conditions for the fermions 
  in the Kaluza-Klein reduction  are chosen to be periodic.  
  
In the case of the bubble of nothing, the background spacetime is simply connected 
and asymptotically approaches a cylinder, where as we described earlier, one must
impose that the fermions be antiperiodic as one goes around the extra-dimensional
circle. It therefore follows that supersymmetry must be broken in the asymptotic pure KK background
state of the bubble of nothing. This is consistent with the results in 
\cite{Henneaux:1984ei,Howe:1995zm,Witten:1994cga,Forste:1996aj} for  purely 
gravitational theories, where it was   shown that covariantly constant spinors do not exist in asymptotically conical  spacetimes.
Indeed, the spacetime of the bubble of nothing  is asymptotically conical with deficit angle of $2\pi$, that is, a cylinder.  
   
   This relation between supersymmetry and the boundary conditions of the fermions can also be understood intuitively by
   looking at the mass spectrum of the KK theory \cite{Scherk:1979zr,Scherk:1978ta}.  When we perform a Kaluza-Klein 
   reduction in the background \eqref{trivialCylinder} but with  boundary conditions other than periodic, some of the fermions 
   that  would be present      
  in the reduced theory acquire masses of the order of the KK scale.
  As a result, 
  supersymmetry is broken in the dimensionally reduced theory.\footnote{See a more detailed description of this point in the Appendix \ref{secondappendix}.}

  This is just another way of stating the result in \cite{Witten:1981gj} that the simple supersymmetric 
vacuum would not be allowed to decay by the formation of a bubble of nothing due to the incompatibility
of the spin structures between the supersymmetric compactification state and the bubble of nothing geometry.

\subsection{Winding compactifications and antiperiodic fermions}

 From our discussion in the previous paragraph, we see that imposing that the background \eqref{trivialCylinder} be 
 supersymmetric introduces a topological obstruction to the  formation of bubbles of nothing.  However, 
 this topological obstruction is not always present in all possible supersymmetric backgrounds. Indeed, it is easy to see that the 
 simple model \eqref{phenomL} admits supersymmetric compactifications demonstrating this. Consider the 
 following bosonic background, which is also of the form of  eqs.~(\ref{KKansatz})-(\ref{KKansatzmetric}),
\be
\phi = \eta\,  \rme^{\rmi \theta}, \qquad A_\mu =e^{-1} \, \delta_{\mu \theta}, \qquad \psi_\mu = \chi = \lambda =0, \qquad ds^2 = -dt^2 + dz^2 + dr^2 + R^2 d\theta^2. 
\label{wilsonCylinder}
\ee

The main  difference with respect to the vacuum considered 
before (\ref{trivialCylinder}) is that the gauge vector has  non-vanishing vacuum expectation value, so that the configuration has  
a $\U(1)$ Wilson line on the compact direction. Furthermore, in  order to satisfy $D_\mu \phi = 0$,   the scalar field also needs to wind around the compact direction. 

As in the previous example, the only non-trivial supersymmetry transformation is the one of the gravitino,
\be
\cD_\mu \epsilon_L = (\pd_\theta +  \ft{\rmi}{2} \kappa^2 \eta^2 ) \epsilon_L =0,  \qquad\Longrightarrow \qquad \epsilon_L(x^\mu) = \rme^{- \rmi \frac{ \kappa^2 \eta^2}{2} \, \theta} \, \epsilon_L^0 .
\label{gravitinoEqCylinder}
\ee
We can see that, provided the parameters of the theory satisfy
\be
\gamma=\kappa^2 \eta^2 = 1 \,,
\label{susy-condition}
\ee
it is possible to find a covariantly constant 
spinor which is consistent with imposing anti-periodic  boundary conditions for the fermions. This is just a special case of the  result found 
in \cite{Becker:1995sp}, showing that covariantly constant spinors may exist in conical spacetimes when they are coupled 
to a $\U(1)$ gauge field.  
Note that the background \eqref{wilsonCylinder} can be consistently interpreted as the asymptotic region of a conical spacetime 
with deficit angle of $2 \pi$, where fermions are necessarily  anti-periodic, and therefore the same mechanism ensures that 
supersymmetry is fully preserved. 

 Intuitively, the relation between the Wilson line, the winding scalar field and supersymmetry breaking can also be understood as before
 by looking at how it affects the KK mass spectrum.  As was argued in \cite{Hosotani:1983xw}, when  the configuration has a 
 Wilson line on the $S^1$, the whole KK mass spectrum of the fermions coupled to it  gets shifted by an amount proportional to 
 the magnitude of the Wilson line. Then,  choosing  conveniently the expectation value of the vector boson and the couplings, 
 it is possible to tune to zero the masses that would be induced by the non-periodic boundary conditions. 
 As a consequence, 
 the field content  left after the reduction is sufficient to form full supermultiplets, as in the case with trivial boundary conditions, 
 and therefore it possible to obtain a low energy theory  invariant under supersymmetry. We describe this arguments in more detail
 in the Appendix \ref{secondappendix}.\\

 Summarizing, the field configuration \eqref{wilsonCylinder}  represents a supersymmetric KK compactification which is consistent 
 with anti-periodic fermions on the circle and, in consequence, it has is no topological protection against decay via   
 nucleation of bubbles of nothing.

\section{The Bubble of nothing geometry }\label{sec:bubblewithstring}

In previous sections we have discussed several compactification
scenarios of our model. We now describe the bubble of nothing
geometries that would represent the decay of these compactifications
to nothing. 

\subsection{Bubble of nothing for pure vacuum solutions}
\label{bubblevacuum}

We start our discussion with a lower dimensional version
of the usual bubble of nothing vacuum solution \cite{Witten:1981gj},
which in 4D is given by the double Wick rotation of the Schwarzschild solution:
\begin{equation}
ds^2 =  \rho^2 \left(-dt^2 + \cosh^2 t ~d\chi^2\right) + \left( 1- {\rho_0\over {\rho}}\right)^{-1}d\rho^2 + \left( 1- {\rho_0\over {\rho}}\right) d\Theta^2~.
\label{rotBH}
\end{equation}
Here $\chi$ is an angular coordinate in   $[-\pi, \pi)$, $\rho \in [\rho_0,\infty)$ is a radial coordinate, and $\Theta$ is a periodic 
variable which runs from $0$ to $2 \pi R$, $R$ being the asymptotic radius of the compact KK dimension. The parameter $\rho_0$ determines 
the size of the bubble  at the time of its formation, $t=0$. This is a vacuum solution of Einstein's equations. 

In order to discuss the geometry of this spacetime it is convenient to introduce a new coordinate system $\{\tau,r,z,\theta\}$, given by
\be 
 t = H_0 \, \tau, \qquad \chi = H_0 \, z, \qquad \Theta = R\,  \theta, 
\ee
where $H_0 = \rho^{-1}_0$, which must be assumed positive for now.  Note that the angular variable $z$ now takes values in $[-\frac{\pi}{H_0}, \frac{\pi}{H_0})$, and the coordinate $\theta$ parametrizing the compact direction runs in $[0,2 \pi)$. The new  radial coordinate $r\in [0,\infty)$ is defined implicitly in terms of the differential equation
\be
\frac{d \rho}{d r} =\sqrt{1 -  {\rho_0\over \rho}}, 
\label{defB}
\ee
and the boundary condition $\rho(0) = \rho_0$ (or $r(\rho_0)=0$).  That is, the position of the bubble is now given by $r=0$.  In this gauge,  the metric takes the form
\begin{equation}
ds^2 =  B(r)^2 \left(-d\tau^2 + \cosh^2( H_0 \, \tau) ~dz^2\right) +dr^2 +  C(r)^2 d\theta^2,
\label{arbitraryR}
\end{equation}
where the metric profile functions $B(r)=H_0\, \rho(r)$ and $C(r)$ are determined implicitly by the expressions
\bea
r(B) &=&  H_0^{-1} \sqrt{(B -1)B} + H_0^{-1}\log\left(\sqrt{B} + \sqrt{B -1}\right), \label{r_B} \nonumber \\
C(r) &=& R \, \sqrt{1 - 1/B(r)}.
\label{implicitR}
\eea

It is worth pointing out that the bubble of nothing geometry has a characteristic property which is an immediate consequence of the definition of $B(r)$ and the equations \eqref{defB} and \eqref{implicitR}: 
\be
B'(r) = H_0 \sqrt{1 - B^{-1}}= H_0 \,  C(r) / R.
\label{BvsC}
\ee
Later on we will use this property to determine when a solution  is an approximate solution of the vacuum Einstein's equations of the bubble of nothing type. More specifically, when the metric functions of our solutions fulfill (approximately) eq.~(\ref{BvsC}), it will mean that the metric configuration resembles that of the pure vacuum bubble of nothing solution.

Taking the limit $r\rightarrow \infty$ of (\ref{arbitraryR}) one identifies the
asymptotics of this solution as   
 $\mathbb{M}_3 \times S^1$ in a coordinate representation similar to the Rindler slicing of Minkowski space,
\be
ds^2 \approx r^2 H_0^2\left(-d\tau^2 + \cosh^2( H_0 \, \tau) ~dz^2\right) +dr^2 +  R^2 d\theta^2.
\label{assympMetric}
\ee
In other words, this geometry asymptotically approaches one of the simple KK compactifications described in
\eqref{trivialCylinder}. One can show that the Euclidean version of \eqref{rotBH}-\eqref{arbitraryR} 
possesses a single negative mode in its spectrum of perturbations, which implies the
existence of an instability for these backgrounds.

\begin{figure}[t] 
\centering
\includegraphics[width=16.0cm]{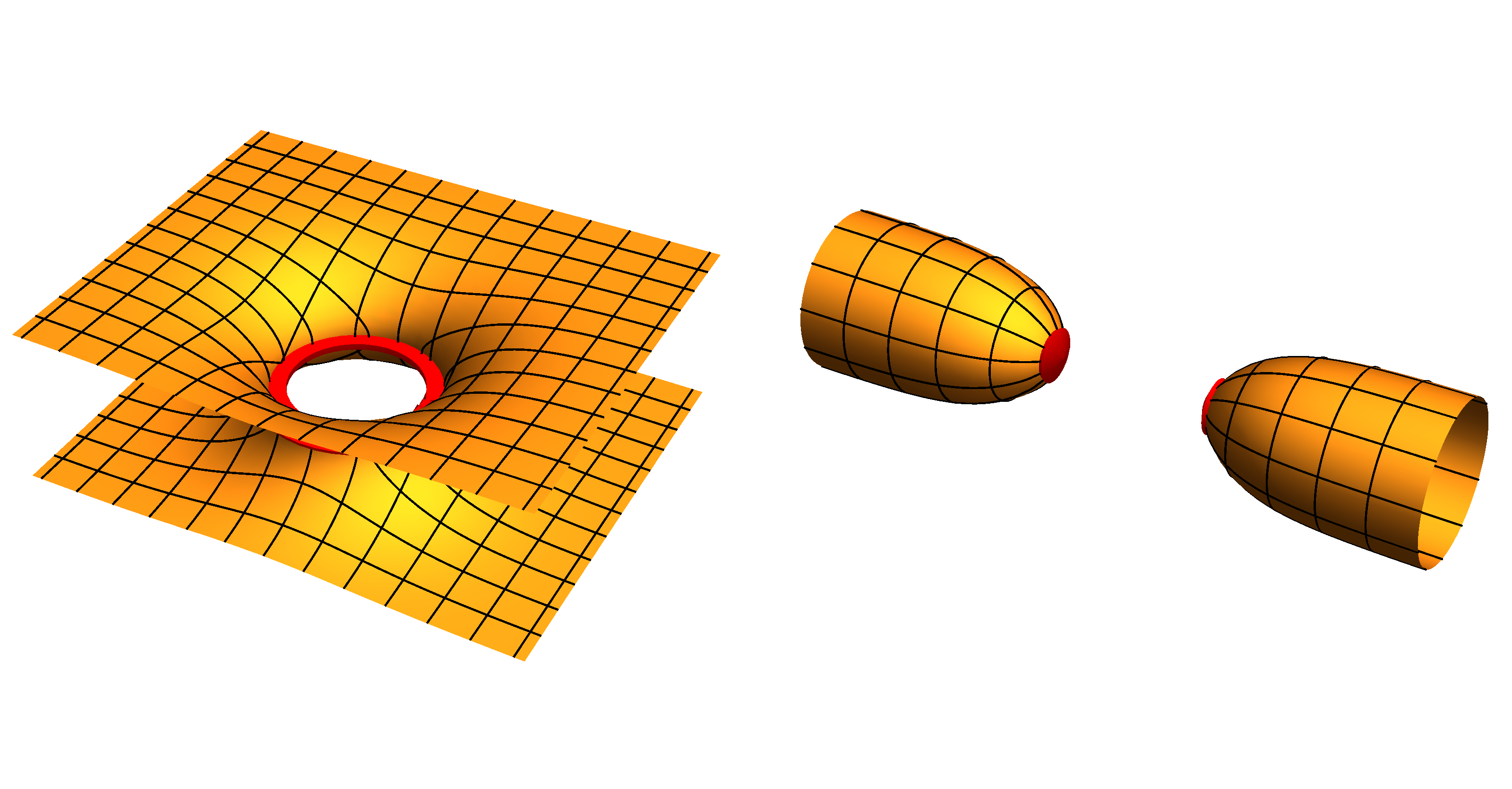}
\caption{{\bf Representation of the bubble of nothing.} Left: This figure shows the two large spatial dimensions
of the geometry with the apparent vertical separation representing the Kaluza-Klein extra dimension. Right: This image
represents the full extra-dimension but only the radial large direction. We use a thick red region in both images to 
represent the vortex/string present at the interior edge of lower-dimensional spacetime, where the extra 
dimension degenerates to zero size, as described in Sec.~\ref{BON-and-STRING}.}
\label{bubble2}
\end{figure}

We can understand the topology of this spacetime by studying its behaviour in the vicinity of the bubble location, at $r \approx 0$  ($B(r ) \approx 1$). In this regime equation \eqref{r_B} reduces to $r(B) \approx 2 H_0^{-1} \sqrt{B -1}$, and the metric has the approximate  form,
\begin{equation}
ds^2 \approx -d\tau^2 + \cosh^2(H_0\, \tau) ~dz^2 +dr^2 +
r^2{{R^2 H_0^2}\over {4}} d\theta^2\,,
\label{nearBubble}
\end{equation}
which shows that the extra dimension degenerates as one approaches $r=0$. 
 Moreover, we can see that in order to avoid any conical
singularity at $r=0$, the radius $R$ of the extra dimension must satisfy the relation $R = 2 H_0^{-1}$. This means that the transverse directions to the
bubble form a kind of smooth cigar geometry that approaches a cylinder
 of fixed radius at large distances from the tip at $r=0$ (see figure~\ref{bubble2}).  
As a consequence, any loop wrapping the extra dimension can be shrunk to nothing
if we take it to the tip of the cigar, so indeed the spacetime is simply connected,
which enforces that the fermions be anti-periodic along the extra dimension. Therefore
one can consider this as the appropriate geometry of the instanton solution
that describes the decay of a non-supersymmetric KK configuration with 
anti-periodic fermions.

Note that the solution \eqref{arbitraryR} depends on a single parameter, $H_0$, which gives us the Hubble scale of the two-dimensional de Sitter 
slice of the spacetime that corresponds to the surface of the bubble,  as well as the initial radius of the
  bubble, $H_0^{-1}=R/2$. Physically, this
scale parametrizes the deviation of the geometry from Euclidean flat space near the $r \rightarrow 0$ ($B(r) \rightarrow 1$) region.
At the same time, it also encodes the information about the size of the compactified extra dimension in the asymptotic limit $r\to \infty$ ($B(r) \to \infty$)  
through the relation $R= 2 H_0^{-1}$.

In summary, this solution describes an asymptotically flat 3D spacetime with a compact fourth dimension of radius $R = 2 H_0^{-1}$.
 In the $r \rightarrow 0$  limit, the extra dimension pinches
off at a ring, which bounds the excised hole in the two large spatial dimensions, and which expands, eating all of future spacetime. 
This ring/hole is shown in figure~\ref{bubble2}, left.  The fact that in the full 4D geometry, the bubble is a co-dimension two submanifold 
is apparent in the right half of figure~\ref{bubble2}.

\subsection{The Bubble of nothing in winding compactifications}
\label{BON-and-STRING}

As we described in previous sections our model allows for
compactifications where the scalar field $\phi$ winds around the extra
dimension. Moreover, the configuration also involves a Wilson line for
the vector field. Putting these two features together along the
extra-dimension does not augment the energy-momentum tensor. This means that there is no
backreaction on the metric due to the presence of these new ingredients, and the solution
is still given by the pure KK
compactification $\mathbb{M}_3 \times S^1$. However, this background cannot have the same bubble decay channel 
as in the absence of these winding modes. The reason is that if we imagine the
geometry of the vacuum bubble of nothing with the Wilson line wrapping the
extra dimension asymptotically, it is clear that since the spacetime is now
simply connected, one finds some magnetic field flux in the vicinity of the
tip of this cigar geometry (see figure~\ref{bubble2}). This means that the analogous
bubble of nothing should be dressed with this flux. Here we propose that
there is a simple configuration that meets these requirements: 
situating an Abelian-Higgs vortex at the tip of the bubble of nothing geometry.
This vortex has all the appropriate charges to match the asymptotic requirements
of our winding background compactification.

Instanton solutions similar to the one discussed here have been presented in the
literature in the context of flux compactifications 
\cite{BlancoPillado:2010df,BlancoPillado:2010et,BlancoPillado:2011me}.\footnote{See \cite{Yang:2009wz,Brown:2010mf} for 
a different approach to bubbles of nothing in this field theory context. See also \cite{deAlwis:2013gka} for some discussion of bubbles of nothing
in models of flux compactification in the String Theory context.}
On the other hand
there is an important difference between those models and the one we study here.
Our winding compactification is not a flux compactification.  (Indeed we have not
induced any potential for the size of the internal dimension.)   Even though
the presence of the vortex is dictated by the boundary conditions at infinity,
unlike a global vortex, there are no long range interactions, so the vortices' effects 
are much more localized.

There are several ways to justify our proposed dressing for the instanton. Having identified the necessity of this
Abelian-Higgs vortex (cosmic string) on the geometry we would like to convince ourselves that
wrapping the string around the ring at the tip of the vacuum bubble of nothing geometry is, in fact, the 
correct configuration for the string in this background. In order to do that, we will first assume that the
 Abelian-Higgs vortex is accurately described by the Nambu-Goto (NG) equations of motion, and that it does not
 distort the background in a significant way.  In other  words, that one can take the string to be 
a probe in the background of a bubble of nothing geometry. Taking into account these approximations 
one can then easily identify a solution of the NG equations 
of motion of a string sitting at the tip of the bubble of nothing geometry where the circle extra dimension
shrinks to zero size. It is perhaps easier to understand this in the Euclidean version of the
solution where the string worldsheet is then wrapping the minimal surface sphere at the tip of the cigar 
geometry. Intuitively it is clear that in this Euclidean geometry there is no other place where this string can go. The Lorentzian continuation
of this solution represents a string being stretched by the de Sitter expansion of the bubble
of nothing that is eating up the spacetime.

We can now estimate what the effect of this string is on the bubble of nothing geometry, 
drawing from our experience on cosmic string spacetimes.  Assuming a low tension for
the string compared to the Planck scale, we can expect that the only effect on the background
would be to introduce a deficit angle on the space transverse to the string, similar to what happens
for a cosmic string in flat space \cite{VilenkinShellard}.

One can introduce such a deficit angle on the metric by changing the value of $H_0$ in \eqref{arbitraryR} to make
it depend on the tension of the string, and at the same time keeping fixed  the radius of the extra dimension $R$ to be  
 what we had before. The last condition ensures that at large distances from the bubble, $r \to \infty$, the 
 spacetime still asymptotes to a KK geometry $\mathbb{M}_3 \times S^1$ with  radius $R$ for the extra dimension, as 
 in \eqref{assympMetric}. More specifically we should take
\be
2 H_0^{-1} =  {{R} \over {1- \frac{\Delta_W}{2 \pi}}} = {{R} \over {1- \frac{\mu \kappa^2}{2 \pi  }}} = {{R} \over {1- 4 G \mu}},
\label{deficit}
\ee
 where $\Delta_W$ is the {\em local\/}\footnote{A local deficit angle is measured at an infinitesimal distance from the Nambu-Goto string.  Farther from the string, $\Delta_W$ is
 the deficit angle removed from Witten's smooth bubble geometry.} 
 deficit angle induced by the string of tension $\mu$.  (Remember that in our notation $\kappa^2 = M_P^{-2} = 8 \pi G$.)
We can now repeat the same calculation that we had done before to obtain the form of the metric in the limit $r \rightarrow 0$. After 
substituting  the previous expression in \eqref{nearBubble} we find
\begin{equation}
ds^2\approx -d\tau^2 + \cosh^2 \left( \frac{2(1- 4 G \mu)}{R} \, \tau  \right) ~dz^2 +dr^2 + (1- 4 G\mu)^2 \, r^2  d\theta^2\,.
\end{equation}
This solution has now a conical singularity at $r\rightarrow 0$ signaling the presence
of the string of tension $\mu$ at that point. Furthermore the circumference of the initial bubble ring, which can be read from the periodicity 
of $z\in [-\frac{\pi}{H_0}, \frac{\pi}{H_0})$,  is now modified
by the string, becoming slightly bigger than before: $H_0^{-1} > R/2$.  We will refer to the vacuum geometry characterized by $H_0^{-1} \neq R/2$ as
a {\em deformed\/} bubble of nothing \footnote{The thermodynamic properties of the euclidean version of this solution are discussed in \cite{Englert:1995mc,Englert:1995je}.}.

These solutions describe the most important modifications of the geometry
for the bubble of nothing instantons in our model with the winding fields
around the extra-dimension. In particular they describe a very interesting property
of the model in the limit of $4G\mu \rightarrow 1$.  
In this {\em critically deformed\/} case, one sees that the bubble size for our instanton becomes infinite, $H_0^{-1} \to \infty$.
In other words,
the string world-sheet becomes flat and the transverse space to all these
solutions corresponds to a cigar-like static geometry. This static, infinite (string wrapped) bubble signals a 
complete suppression of the tunneling process to nothing, exactly in the same way as what 
happens in the usual Coleman - de Lucia \cite{Coleman:1980aw} transition. The Euclidean action in this
case diverges, and the decay rate to nothing vanishes much in the same way as 
it occurs in field theory models without extra dimensions.

It is important to note that these configurations have been found using the thin wall approximation,
and it is not clear if the suppression will survive in 
the full field theory description of our model.  In the following sections we will test all these 
ideas by looking at the smooth numerical solutions of the bubbles of nothing
within the Abelian-Higgs model.

\section{Bubble of nothing in the Abelian-Higgs model}
\label{sec:AHBON}

Previous arguments suggest the existence of solutions describing the decay of
a compactified space via the formation of a bubble of nothing 
where the bubble is {\it dressed} with a cosmic string. In this section we would like to explore
the existence of these solutions in the Abelian-Higgs model where the
cosmic string will be represented by
a smooth vortex. Our starting point is the action,
\begin{equation}
S= \int{d^4x \sqrt{-g} \left({{1}\over {2\kappa^2}}R - |D_{\mu} \phi|^2- {1\over 4}F_{\mu \nu}  F^{\mu \nu}-  {{\beta e^2}\over{2}} (\eta^2 - \phi \bar \phi )^2\right)},
\end{equation}
where we have introduced the deformation parameter $\beta$. Note that this action only coincides with the bosonic sector of the supergravity 
model presented in section \ref{sec:model} for $\beta =1$, and thus this parameter  determines an explicit breaking of supersymmetry 
for values of $\beta \neq1$.

We will look for solutions with a generalized bubble of nothing ansatz for the metric, namely solutions of the form \eqref{arbitraryR}, with the 
profile functions $B(r)$, $C(r)$ and the parameter $H_0$ yet to be determined.
As in the previous section  the induced metric on the bubble wall (and the vortex) is a 2-dimensional de Sitter space with Hubble 
parameter $H_0$, and the initial bubble radius at $\tau=0$ is given by $H_0^{-1}$.  The limiting behaviour of $C(r)$ for large values of 
the radial coordinate $r$ fixes the asymptotic size of the compact dimension via the relation $R= \lim_{r\to\infty}C(r)$.

We will consider the configuration for a vortex of unit winding number for the matter fields, which is of  the form  \cite{VilenkinShellard}, 
\begin{equation}
\phi(r) = f(r) e^{i  \theta}, \qquad A_{\mu} =  \ft 1e (1-a(r)) \delta_{\mu \theta}.
\label{matterfield}
\end{equation}

In order to simplify the notation, we can redefine fields and lengths by the following rescalings,
\be
f \to \eta f,  \qquad \tau \to l_g \tau, \qquad r \to l_g r, \qquad z \to l_g z,  \qquad C \to l_g C, \qquad H_0 \to l_g^{-1} H_0,
\label{rescalings}
\ee
where we are using the length scale $l_g \equiv \ft 1{\eta e}$, corresponding to the vector core thickness. Note that all the 
coordinates $\{\tau,r,z,\theta\}$ and the the parameters $R$ and $H_0$ are now dimensionless. Using this ansatz we arrive at the matter field equations 
\be\label{eq:mattereom}
{{(B^2 C f')'}\over {B^2 C}} - {{a^2 f}\over{C^2}} + \beta  (1- f^2) f=0,\qquad \frac{C}{B^2}\left({{B^2 a'}\over {C}}\right)' -2  f^2a  =0.
\ee
Furthermore the $t$-$t$ component and the $\theta$-$\theta$ component of the Einstein's equations read (the $r$-$r$ component is a constraint) 
\bea
 \frac{(CBB')'}{B^2C} &=& \gamma \left(\frac{ a'^2}{2 C^2} - \frac{\beta}{2} (1-f^2)^2\right) + \frac{H_0^2}{ B^2}\,,\nonumber\\
\frac{(B^2C')'}{B^2 C} &=& - \gamma \left(\frac{a'^2}{2 C^2} + \frac{2  a^2 f^2}{C^2} + \frac{\beta}{2} (1-f^2)^2 \right)\,,\label{eq:Ceom}
\eea
where we have made use of the dimensionless parameter $\gamma= \kappa^2 \eta^2$ introduced earlier which, as we see from these 
equations, determines the gravitational coupling of the string. 
 
\subsection{Compactificatified vacuum states}

Using the ansatz given above it is easy to show that the following configuration solves the equations of motion
for  arbitrary values of $\beta$ and $\gamma$,
\be
f(r)= 1,\qquad a(r) =0 , \qquad B(r)=H_0 \, r,\qquad C(r)=R~.
\ee

Putting the dimensionful constants back in, we can see that this is nothing more than 
our original $\mathbb{M}_3 \times S_1$  background given in the same gauge as in eq.~(\ref{assympMetric}),
\be
\phi = \eta\,  \rme^{\rmi \theta}, \qquad A_\mu =e^{-1} \delta_{\mu \theta}, \qquad ds^2 =  r^2 H_0^2\left(-d\tau^2 + \cosh^2 ( H_0 \, \tau) ~dz^2\right) + dr^2 + R^2 d\theta^2. 
\label{compactification-solution}
\ee

We note that fixing the values of $\beta$ and $\gamma$ we completely specify the theory to be considered but we
still have the freedom to set the radius of the compactified space, $R$, to any value. This is just
a reflection of the fact that the radius of the extra dimension is a flat direction in the moduli
space of the compactified theory. The constant $H_0$ is also left undetermined in this background, but here it has no physical meaning, 
it just signals a coordinate freedom associated to the Rindler slicing we are using to parametrize the $\mathbb{M}_3$.

\subsection{Boundary conditions for the bubble solutions}
\label{sec:bc}

In order for the spacetime of the bubble of nothing to be everywhere regular, 
 the metric profile functions must satisfy the following conditions on the bubble, that is, at $r=0$ (see appendix \ref{Numerics}):
 \be
 C(0)=0, \qquad  C'(0) = 1, \qquad B'(0)=0. 
\label{initialConditions}
 \ee
We will also require the gauge condition $B(0)=1$, in order for the metric to have the form \eqref{nearBubble} in the limit $r \to 0$.
These conditions mean that the geometry near the bubble approaches that of  $dS_2\times \mathbb{R}^2$, where the two factors 
represent  the intrinsic de Sitter geometry on the bubble  and the smooth end of the compact dimension, respectively. In other 
words, for $r \approx 0$ we have
 \be
ds^2 \approx -d\tau^2 + \cosh^2(H_0\, \tau) ~dz^2 +dr^2 +
r^2 d\theta^2.
\label{coreGeometry}
\ee
Note that, since $z \in [-\frac{\pi}{H_0},\frac{\pi}{H_0})$, the limit $H_0\to 0$ of this geometry is locally that of 4-dimensional Minkowski space.

For the matter field configuration to be  regular at the bubble location we must also require that 
\be
f(0)=0, \qquad  \qquad \qquad a(0)=1.\label{boundaryConditions1} 
\ee

Solutions   representing a bubble of nothing in a winding compactification should approach a KK vacuum of the form 
\eqref{compactification-solution} asymptotically, therefore the profile functions must have the following asymptotic behaviour for $r \to \infty$:
\be
\lim_{r\rightarrow \infty} f(r) = 1, \qquad \qquad \lim_{r\rightarrow \infty} a(r) = 0, \qquad \qquad  \lim_{r\rightarrow \infty} C(r) = R.
\label{boundaryConditions}
\ee

Note that, a priori, the parameter $H_0$ appears to be  unfixed  by the boundary conditions. However once we demand the profile 
functions to meet all the boundary conditions specified above, there is a unique value of $H_0$ which is compatible with them and 
the fields equations.

 In summary, we will obtain the parameter $H_0$, together with the form of the profile functions of the metric and matter fields as a result of numerically 
 solving eqs.~\eqref{eq:mattereom} and \eqref{eq:Ceom}, subject to the conditions \eqref{initialConditions}, \eqref{boundaryConditions1} and 
 \eqref{boundaryConditions}.

\section{Comparing  numerical results with the  thin wall approximation}
\label{sec:thinwall}

We have shown in previous sections that one can obtain a supersymmetric compactification
by specifying the condition,
\be
\beta =1\,, \qquad\gamma =1\,.
\ee
In the following we will consider different values of these parameters and their approach to the
supersymmetric limit. In this section we will consider the regime of parameter space where the vortex 
 size is much smaller than the compactification radius $R$ and the initial bubble size $H_0^{-1}$. This 
 is the situation where we expect the thin  wall approximation discussed in section \ref{BON-and-STRING} 
 to be an accurate description.

\subsection{Non-supersymmetric compactification: the $\gamma \neq 1$ case}

We start our investigation by looking at solutions where $\beta=1$ and the gravitational coupling of the vortex is small, $\gamma \ll 1$. 
 It is clear
that in this case, the compactification would break supersymmetry spontaneously since
the condition in eq.~(\ref{susy-condition}) will not be satisfied. 

Keeping $\beta=1$, the vortex solution is special in the sense that 
the scalar and magnetic cores are of the same size, which is unity in our current units\footnote
{In flat space, the value of $\beta$ determines if we are in the type-I ($\beta<1$) or 
type-II regime ($\beta>1$).  $\beta=1$ is the Bogomolnyi limit. For $\beta=1$ the 
vortex core radius is given by $l_g$ without the rescalings in eq.~\eqref{rescalings}.
} 
\cite{VilenkinShellard}. On the other hand, we still have the
freedom to fix the size of the compact dimension, so it makes sense to start
our investigation in the regime where there is a clear separation of scales
between the size of the vortex and the size of the extra dimension, $R\gg 1$.  It is
in these cases that we expect a bubble of nothing solution that is very similar
to Witten's pure gravity solution. Furthermore, small variations from this solution 
should be well captured in this regime of parameters by our analysis within the 
thin wall approximation.

\begin{figure}[t]
\centering \makebox[\textwidth][c]
{ \hspace{0cm}\includegraphics[width=0.62\textwidth]{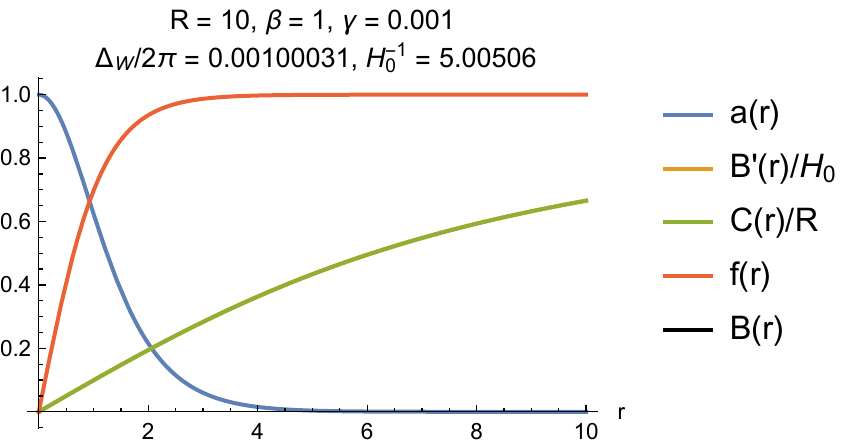}\hspace{.1cm}\includegraphics[width=0.44\textwidth]{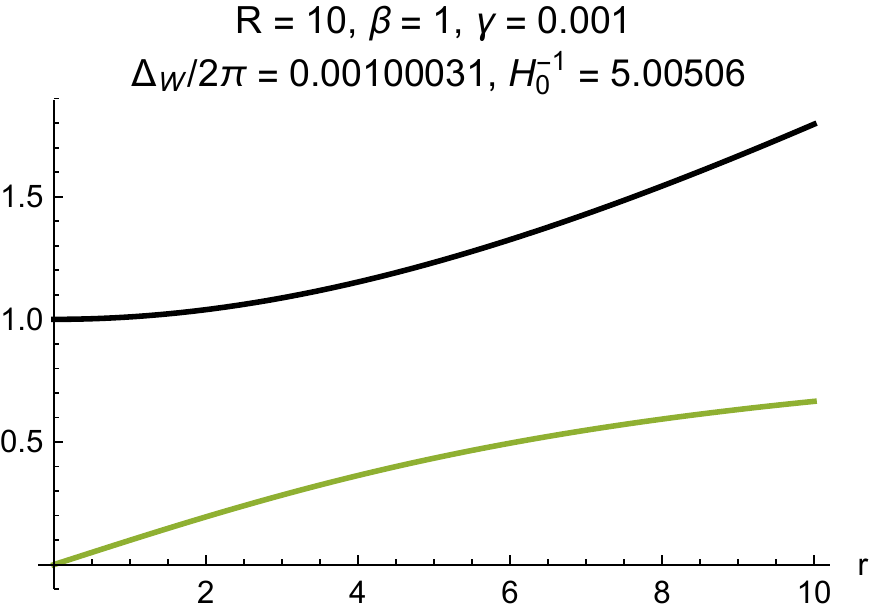}}
\caption{Bubble of nothing in the Abelian-Higgs model with $\beta=1$, $\gamma = 10^{-3}$ and 
asymptotic radius of the compact dimension $R=10$.  Note that since $R\gg 1$, the vortex is 
much thinner than the background curvature, and so we can compare the bubble deficit angle 
$\Delta_W \approx 2 \pi\, 10^{-3}$ with the flat space deficit angle $\Delta = 2 \pi\, 10^{-3}$ in 
eq.~(\ref{eq:susyDelta}). This is a very light vortex with tension $4 G\mu = 10^{-3}$, so 
backreaction is almost negligible. In the left panel, the matter field profiles are shown, together with 
the metric fields $B'(r)/H_0$ and $C(r)/R$. These last two are on top of each other, inferring that we are in 
a bubble of nothing configuration, as given by (\ref{BvsC}). This can be corroborated by the profiles 
of $B(r)$ and $C(r)$ in the  right panel.}
\label{bubble1}
\end{figure} 

In figure~\ref{bubble1} we present a numerical solution that corresponds to a bubble of 
nothing in such a regime, where we have fixed $\gamma=10^{-3}$ and $R =10$. 
We relegate to Appendix \ref{Numerics} the
detailed explanations of the numerical procedure we use in order to find these solutions. Using 
those techniques and given the values of the parameters $(\beta, \gamma, R)$, we are able 
to compute both the values of the initial size of the bubble $H_0^{-1}$, and the vortex induced deficit angle  $\Delta_W$
that one can infer from the asymptotic vacuum solution.

It can be shown that the effect of the string vortex
on the bubble of nothing is negligible and the profile functions $B(r)$ and $C(r)$ resemble 
almost exactly the form given by Witten's bubble configuration \eqref{arbitraryR} and \eqref{implicitR}. The relation (\ref{BvsC}) is satisfied everywhere, 
as it can be checked in figure~\ref{bubble1}, where the line representing $B'(r)/H_0$ remains hidden by the one associated to $C(r) /R$.   As mentioned before,
this property is characteristic of the  pure vacuum bubbles of nothing. Another way to quantify this is by looking at the ratio
between the values of   $R$  and $H_0^{-1}$, which in this case is very close to $2$ as in the original
bubble of nothing.  In the right panel of figure~\ref{bubble1} we have displayed the asymptotic regime of the profile functions $B(r)$ and $C(r)$, 
showing that  the spacetime geometry approaches $\mathbb{M}_3 \times S^1$ in the limit $r\to \infty$.

 The most notable difference with \eqref{arbitraryR} and \eqref{implicitR} 
is given for the asymptotic presence of a deficit angle on this solution, which in our numerical solution is  
$\Delta_W \approx 2 \pi \, 10^{-3}$. We can now compare
this value with the deficit angle for the vortex in an asymptotic conical 
space for the same values of the parameters $\beta, \gamma$. In our case, we
can use the result for the case for a vortex of a single unit of flux  \cite{VilenkinShellard},
\be\label{eq:susyDelta}
\left.\Delta(\gamma)\right|_{\beta=1} = 8\pi G ~\left.\mu(\gamma)\right|_{\beta=1}  = 2 \pi \gamma ~ =2 \pi \, 10^{-3}.
\ee

We see that the result obtained from the numerical calculation agrees perfectly with
the analytic results described earlier. Furthermore, the profile of the matter  fields, $f(r)$ and $a(r)$, are not very much affected by
the existence of the bubble. This is not surprising, since the radius of the bubble
$H_0^{-1}\approx R/2= 5$ is large compared to the size of the defect.

\subsection{Explicit supersymmetry breaking: the $\beta \neq 1$ case}

\begin{figure}[t]
\centering \hspace{-.005cm}\includegraphics[width=0.8\textwidth]{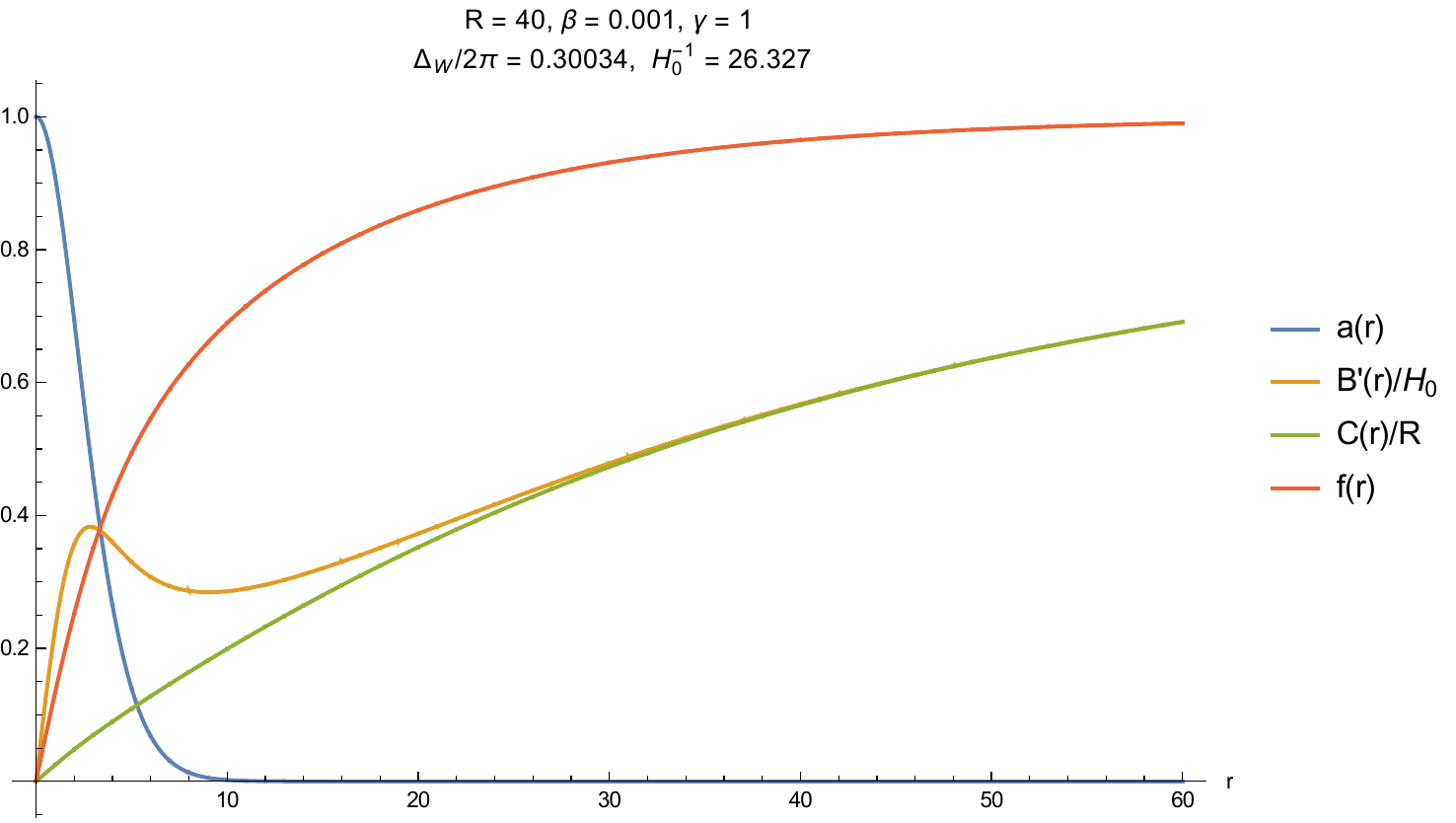}
\caption{Bubble of nothing in the Abelian-Higgs model with explicit breaking of supersymmetry, $\beta= 0.001$, strong 
gravitational coupling $\gamma=1$, and large asymptotic radius of the compact direction $R= 40$. Note that the 
relation (\ref{BvsC}) is satisfied far from the core.}
  \label{bubble-low-beta}
\end{figure}

We can also explore the regime where we set $\gamma=1$ and vary $\beta$. In this case,
the original theory breaks supersymmetry explicitly, so we expect the existence of bubbles
of nothing in this case as well.  The  thickness of the scalar vortex gets bigger when decreasing the value of $\beta<1$,  
so in order to check the validity of the thin wall approximation we should also consider 
large values of the extra dimensional space and integrate the equations to larger distances
from the core. 

We show in figure~\ref{bubble-low-beta} a solution for $\beta=0.001$, and 
 $R=40$. This solution agrees well with the thin wall approximation of a conical defect produced by
the analogue vortex in an asymptotically locally flat spacetime that is placed on a
bubble of nothing geometry. For instance we see that far from the vortex core ($r\gg1$) the metric 
profile functions approach those of the standard bubble of nothing \eqref{arbitraryR} and 
\eqref{implicitR}, and satisfy the characteristic relation \eqref{BvsC}. 

Although  the string tension decreases for smaller values of the
 parameter $\beta$ the dependence is  logarithmic \cite{VilenkinShellard}, which means that it is 
 dominated by the large value of $\gamma$ in our case. This explains why there is still a quite important
backreaction on the value of $H_0^{-1}\approx 26 > R/2=20$ compared to the pure bubble of nothing
geometry in our example. Note also that the equation \eqref{eq:susyDelta} for the deficit angle of the 
vortex in an asymptotically conical spacetime is only valid for $\beta=1$, and therefore it cannot be used 
in the present case to predict the approximate value of  $\Delta_W$. Nevertheless, we have checked that
the obtained $\Delta_W$ agrees well with the analogue value in a conical spacetime for the same
parameters.

These solutions and their agreement with the thin wall approximation described in section \ref{BON-and-STRING}
validates our numerical techniques and demonstrates explicitly the existence of these decay channels
in  models with broken supersymmetry.

\subsection{Supersymmetric limit: Approaching the critical bubble}
\label{susyThinWall}

\begin{figure}[t]
\centering\makebox[\textwidth][c]
{ \hspace{0cm}\includegraphics[width=0.62\textwidth]{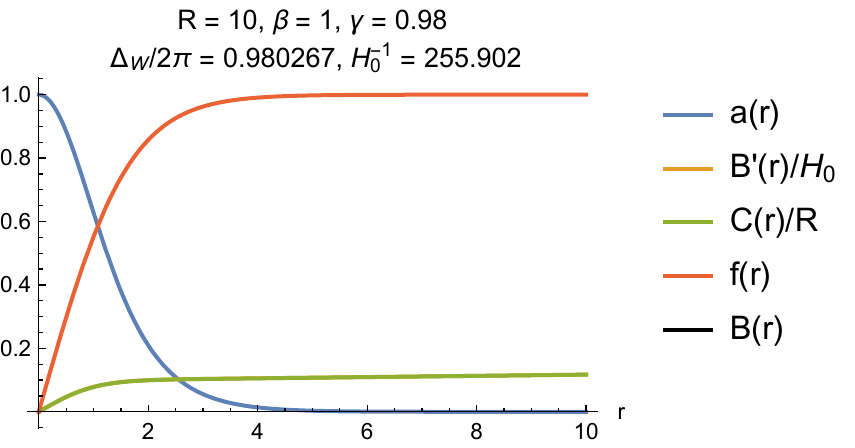}\hspace{0.1cm}\includegraphics[width=0.44\textwidth]{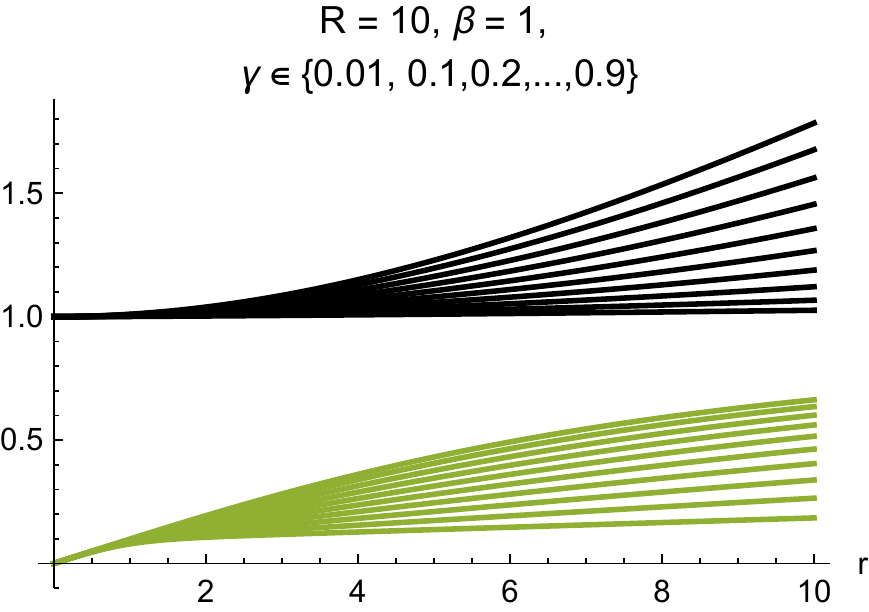}}
\caption{In the left panel we plot a bubble of nothing configuration near the supersymmetric limit ($\gamma=0.98$).  
Note that although the matter fields rapidly approach their asymptotic (vacuum) values, in the region plotted the extra dimension 
has size $C(r) \approx 1$, and it only very slowly approaches its asymptotic value of $10$.  The right plot shows a sequence of 
bubble of nothing geometries for increasing values of $\gamma$, fixed $\beta=1$, and $R=10$. From top to bottom  
$\gamma =\{0.01, 0.1, 0.2, 0.3,\ldots, 0.9 \}$, showing that for higher values of $\gamma$ there is a larger region 
($1 \lesssim r \lesssim H_0^{-1}$) where the metric profile functions $B\approx 1$ and interestingly, $C \approx 1$, 
regardless of the value of $R$.}
\label{bubble-sequence}
\end{figure}

Taking as our initial condition the solutions found previously, we would like to find what happens
as one approaches the supersymmetric compactification limit where $\beta=1$ and $\gamma=1$. As in the previous subsections, 
we will restrict our attention  to the case where the asymptotic compactification radius is large compared to the vortex width, 
$R \gg 1$, so that the predictions of the thin wall approximation are applicable. Nevertheless, due to the large value of the 
gravitational coupling, $\gamma$, the  backreaction of the vortex on the geometry is expected to be large, and thus to induce 
significant deviations from Witten's bubble of nothing  given by equations \eqref{arbitraryR} and \eqref{implicitR}. We show a 
sequence of the numerical solutions in the right panel of figure~\ref{bubble-sequence} with the parameter $\beta$ fixed to unity, 
the asymptotic size of the compact dimension set to $R = 10$,  and  $\gamma$ varying in the range $[0.01,0.9]$.

Moving slowly in the parameter space towards this supersymmetric limit, we see the appearance of two different vacuum regions. 
On the one hand, for $0\leq r \ll H_0^{-1}$, the geometry of the space adjacent to the vortex becomes increasingly similar to  the 
background  of a string in an asymptotically  conical spacetime  with critical tension, i.e. with deficit angle $\Delta\approx 2 \pi$ \cite{VilenkinShellard}.  
Outside the vortex core, $1 \ll r \ll H_0^{-1}$, this vacuum geometry resembles  a static cylinder where the radius of the extra 
dimension is constant and equal to the {\em vortex size\/}, which in our units is $C(r) = 1$.  This region of space is displayed in 
figure~\ref{bubble-sequence}. At large distances from the vortex core, $r \gtrsim H_0^{-1}$,  the spacetime begins to resemble the 
pure compactification, and matches the boundary condition $\lim_{r\to \infty}C(r) =R=10$.  It is instructive to view the different 
behaviors over a large range of $r$, and so we will plot the profile functions over several decades in $r$  in later sections 
(e.g., right panel of figure~\ref{bubble-C-lt-1}), showing how the solutions interpolate between these two regimes.

As one approaches the supersymmetric case, $(\gamma=1,\beta=1)$,   the value of $H_0$ decreases quickly, becoming zero in 
that limit.  Recall that $H_0$ is the Hubble scale of the induced de Sitter geometry on the bubble surface.  The vanishing of this 
parameter thus indicates the arrival at a flat bubble geometry.  The values of $H_0$ obtained numerically are in very good agreement 
with the predictions from the thin wall approximation, i.e., with   equations   \eqref{deficit} and \eqref{eq:susyDelta}.  
Note that the circumference of the bubble at the moment of nucleation is $2\pi H_0^{-1}$.

In summary, in the supersymmetric limit the bubble becomes flat and infinitely large. Moreover, as one approaches this limit, 
the geometry of the transverse directions ever more slowly interpolates between two regimes: 
\begin{itemize}
\item  $1\lesssim r\ll H_0^{-1}$:  outside (but near) the vortex core, the geometry resembles a cylinder of approximately 
constant radius $C(r)\approx 1$. This region becomes infinitely large in the supersymmetric limit where $H_0\to 0$.
\item $H_0^{-1} \ll r$: far from the vortex core, the compact dimension approaches the asymptotic radius $C(r) \approx R$.  
In the supersymmetric limit ($H_0 \to 0$), this asymptotic regime occurs at an infinite distance from the bubble, assuming $R\neq 1$.
\end{itemize}

We show in the left panel of  figure~\ref{bubble-sequence}
the profiles obtained close to the limiting case ($\gamma=0.98$).  In this example the initial bubble size is given by 
$H_0^{-1} \approx 256$, which still in  good agreement with the prediction obtained from the thin wall approximation 
\be
\left.H_0^{-1}\right|_{\rm tw}= {{R}\over{2(1- 4 G\mu)}} \approx 250,
\ee
where we have used  \eqref{deficit} and \eqref{eq:susyDelta}. 
  This is important for the calculation of the decay rate of the 
compactified spacetime, since an infinite bubble would give rise to an infinite action for the instanton and therefore
a total suppression of the tunneling transition. This is exactly what happens
in the usual Coleman-De Luccia suppression mechanism.

Although not plotted here, we have repeated the numerical calculations following different paths towards the 
supersymmetric limit in the $(\beta, \gamma)$ parameter space while keeping fixed the value of $R$, and in 
it is interesting to note that the behaviour we just discussed is independent of the path.

\section{Numerical results away from the thin wall regime}
\label{sec:thickwall}

As we noted earlier, all possible values of the asymptotic KK radius $R$ are allowed for each point in the 
parameter space $(\beta, \gamma)$. We expect
that there will be instanton solutions representing the decay to a bubble of nothing for all these values.
We have argued that these instantons should involve the presence of a vortex
in their geometry that should fit inside the bubble of nothing solution with the correct 
asymptotics. In the previous sections we have shown explicitly that this is possible
in our smooth Abelian-Higgs model when there is a clear separation of scales between
the compactification size and the vortex core size $R \gg 1$.

Here we would like to numerically explore
what happens when we are not in the above regime, in other words, when we are well outside
of the region of validity of the thin wall approximation. 
This is not just a technical curiosity, it is an important point for the conclusions of our paper. 
We are arguing that the supersymmetric
limit of our compactification is protected from the bubble of nothing decay dynamically
by the Coleman-De Luccia suppression mechanism. If this is the case, it should be the same 
for any value of the compactification radius $R$, not only for the situations that are easily described by 
the thin wall approximation. We therefore extend our investigations to some of the 
cases where one can only find the solution by performing the numerical integration 
of the equations of motion.

\subsection{Small compactification radius, $ R\lesssim 1$}

The situation seems more problematic in cases where the
compactified space is smaller than the vortex core. It would seem difficult to find the 
bubble of nothing instanton of the kind that we have been discussing, since there
seems to be no space for the vortex to fit in this geometry. 

\begin{figure}[htbp]
\centering \makebox[\textwidth][c]
{\hspace{2.2cm}\includegraphics[width=0.62\textwidth]{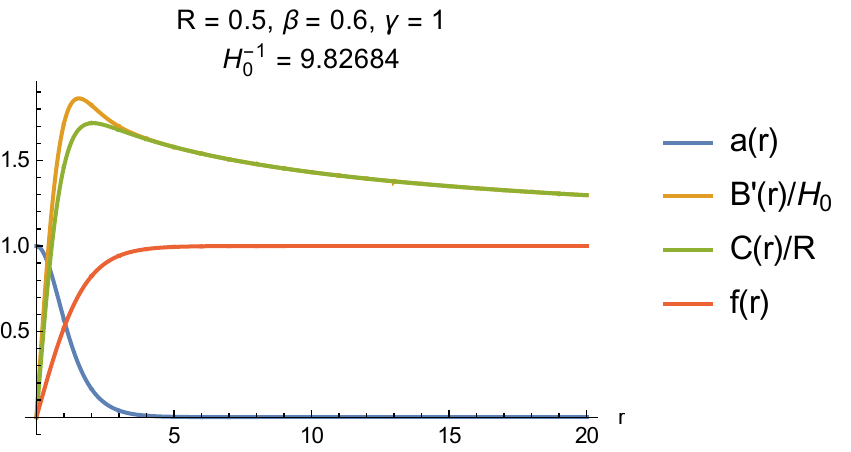}\hspace{.1cm}\includegraphics[width=0.62\textwidth]{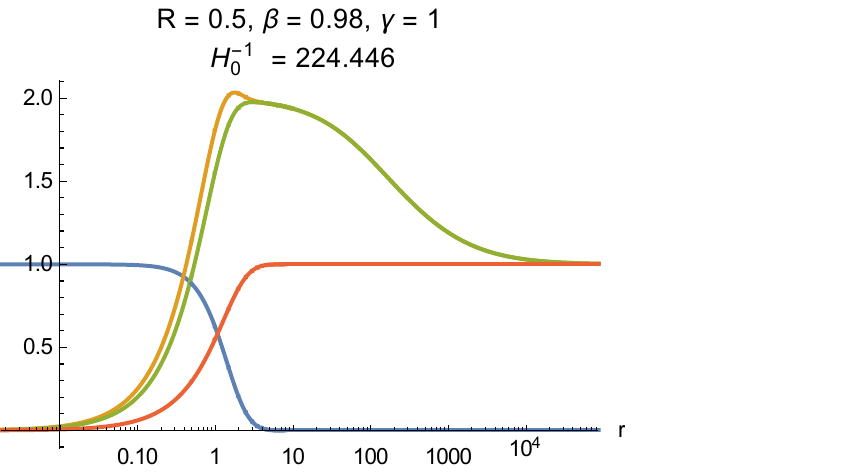}}
\caption{A bubble of nothing for a winding compactification with KK radius $R=0.5$ and gravitational coupling $\gamma = 1$. The
 left panel shows the solution for $\beta=0.6$.  The right panel shows the ($\beta=0.98$) bubble of nothing, close to the supersymmetric case.  
 In this panel the radial distance is arranged on a logarithmic scale so that the figure shows the two regimes: inside/near the core, 
 and asymptotically the compactification boundary conditions.}
\label{bubble-C-lt-1}
\end{figure}

In the left panel of figure~\ref{bubble-C-lt-1}, we present an example of such an instanton for the case of $R = 0.5$,
with supersymmetry breaking parameters $\gamma=1$ and $\beta = 0.6$.    We see that the solution does exist, but its geometry is quite
different from the previous cases. Close to the vortex, the extra dimension is larger than
its asymptotic value. This is due to the presence of the vortex matter fields that force
the extra-dimensional volume to be large enough to hold the vortex. Once the matter fields are settled
near their vacuum values $f(r) = 1$ and $a(r) = 0$, the geometry relaxes (possibly very slowly) to the one imposed by the boundary
condition at infinity.  In particular, we see that the metric profile functions satisfy the characteristic relation \eqref{BvsC}. 
As we discussed above,  the regions where this relation is satisfied  signal that the spacetime metric is an approximate solution of the vacuum 
Einstein's equations.  The interesting point about this configuration is that it settles to a vacuum solution
which corresponds to a Schwarzschild-like solution of the type given by eq.~(\ref{rotBH}) but 
with a negative mass term.  Indeed, when the metric is written in the gauge \eqref{arbitraryR} this means that the metric profile function $C(r)$ has 
the following asymptotic behaviour for $r \gg 1$ ($B(r)\gg1$):
\be
B'(r)\approx  H_0^{-1}>0, \qquad \qquad
C(r) \approx R \, \sqrt{1 + B(r)^{-1}},
\label{negativeM}
\ee
which can be obtained proceeding as in section \ref{bubblevacuum}, but setting $\rho_0<0$.
This explains how the size of the extra dimension, $C(r)$, can be a decreasing 
function of the distance from the core, as figure~\ref{bubble-C-lt-1} shows.

A vacuum solution with this behavior cannot exist on its own, since this would lead to a naked singularity, not a
smooth bubble of nothing geometry. The reason is that a negative mass Schwarzschild
solution does not have a horizon, so both the Lorentzian geometry as well as its analytic continuation would be singular.  
It is only due to the presence of the vortex that
one can cap the geometry, replacing the singularity by the smooth vortex.\footnote
{It would be interesting to investigate these new configurations as regular Euclidean black hole solutions with
negative mass terms.
} 
In this sense, these solutions are clearly not a deformation of the usual vacuum bubble of nothing geometry.


Taking the supersymmetric limit of these solutions, we arrive at the same conclusion as in the previous section.
As one approaches the $\beta=1, \gamma=1$ limit, the size of the bubble, $H_0^{-1}$ diverges,
signaling again the suppression of the decay. We show in the right panel of  figure~\ref{bubble-C-lt-1} an example of such
behaviour for $R=0.5$,  and parameters $(\gamma=1, \beta=0.98)$. In this case the value we obtain for the Hubble 
parameter is $H_0\approx 4 \times 10^{-3}$. We have plotted the solution on a logarithmic axis to see clearly the two regions
in the solution (which we described in section \ref{susyThinWall}), the vortex cylinder close to the tip, $1 \lesssim r \ll H_0^{-1}$, 
and the compactification geometry at large distances, $H_0^{-1}\lesssim r$.  Outside the vortex, this case corresponds to 
an analytically continued Schwarzschild solution with a negative 
mass term.  In particular, we see that for $1 \lesssim r \lesssim  200$ the radius of the compact dimension is given by the 
vortex size $C(r) \approx 1$, or equivalently $C(r)/R \approx 2$, and far away from the core, $ r \gg 200 $, it has {\rm decreased\/} toward the 
asymptotic value $C(r) \to R$.

\subsection{Intermediate regime, $R \sim O(1)$}
\begin{figure}[htbp]
\centering \hspace{-1.cm}\includegraphics[width=1.1\textwidth]{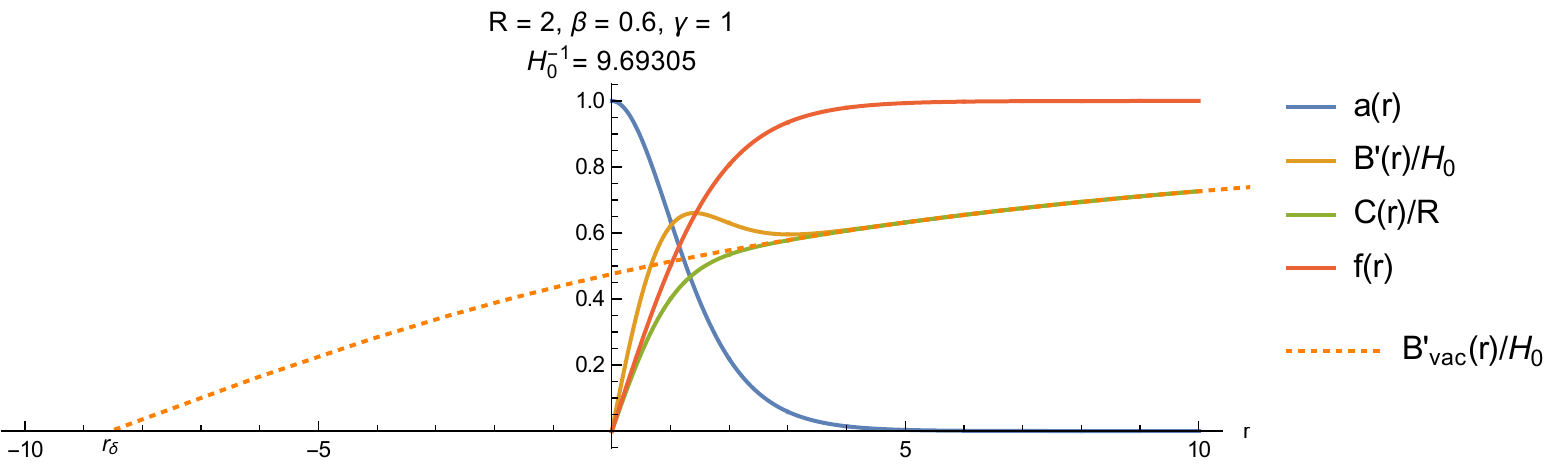}
\caption{A bubble (solid lines) superimposed with the corresponding asymptotic vacuum solution 
$B_{\rm vac}'(r) = H_0 \, C_{\rm vac}(r)/R$ (dotted).   Numerically we also find  $H_0=0.103$ even 
though the asymptotic vacuum solution  would extrapolate 
to a smaller (Nambu-Goto wrapped) bubble with $H_\delta = 0.132$.  We can characterize this ``thick-vortex'' effect as the ratio $H_\delta/H_0 = 1.29$.}
\label{bubble-C-2}
\end{figure}

\begin{figure}[t]\centering \makebox[\textwidth][c]
{\hspace{2.2cm}\includegraphics[width=0.62\textwidth]{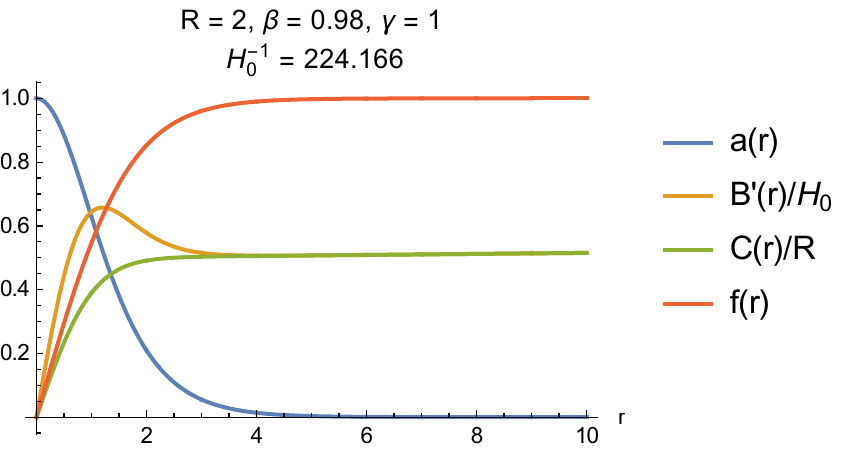}\hspace{.1cm}\includegraphics[width=0.62\textwidth]{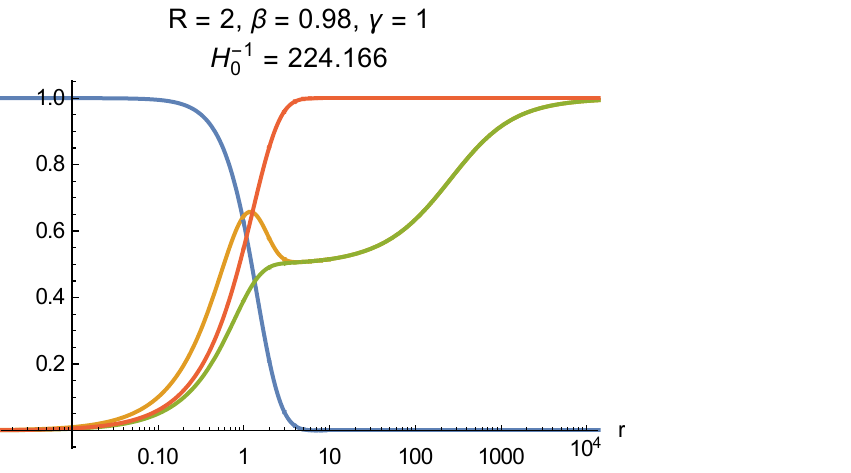}}
\caption{Solutions for  $\gamma=1$ and $R=2$ close to the supersymmetric limit, $\beta = 0.98$.  Note that the initial bubble radius 
$H_0^{-1}\gg1$ is very large.  The behaviour of the profile functions near the core of the vortex, $r \lesssim H_0^{-1}$, are shown in 
the left plot. In the right plot the radial coordinate is represented in logarithmic scale in order to display the behaviour far from the 
vortex core $r \gg H_0^{-1}$.  
\label{limit-bubble-C-2}}
\end{figure}

We can also explore numerically the solutions that interpolate between the extreme
cases discussed earlier, i.e., cases with very small and very large compactification radius. In figure~\ref{bubble-C-2} we 
show a solution with $R = 2$.   We have also represented, with a dashed line, the deformed bubble solution of the vacuum Einstein's equations 
given by \eqref{arbitraryR} and \eqref{implicitR}, which matches the  same asymptotic behaviour of the fully numerical solution.  
As we discussed in section \ref{sec:bubblewithstring}, such a vacuum solution can be completely characterized by the 
boundary condition $R = 2$, as well as the deficit angle $\Delta_W = 5.45$, and  satisfies the relation \eqref{BvsC} \emph{everywhere}. 
Such solutions require a delta-function (Nambu-Goto) source to induce the deficit angle deformation of Witten's bubble, $\Delta_W$.  It should be 
located at a point $r = r_{\delta}<0$ to correctly match the profile functions for $r \to\infty$. Instead, in the fully numerical Abelian-Higgs solution,  
the presence of matter causes a sudden  drop in both $C$ and $B'$, with the bubble appearing at $r=0$, rather than the extrapolated 
value $r = r_\delta$.

Similarly to what we did before, we show in figure~\ref{limit-bubble-C-2} the solution near the supersymmetric limit, $\beta \to 1$. 
The behaviour is similar to the other cases, and in particular we observe that $H_0$ becomes arbitrarily small, implying that 
the suppression persists for all values of the asymptotic  compactification radius $R$.  This intermediate
regime allows us to distinctively see both regions of the deformed bubble solution, the vortex core which resembles
a static vortex solution and the large $r$ geometry that matches the
pure compactification.  In figure~\ref{limit-bubble-C-2}, we display the spacetime region near the vortex core $r \lesssim H_0^{-1}$ on the left panel, and  
 the transition between the vortex core region and the asymptotic geometry for $ r \gtrsim H_0^{-1}$ is shown on the right.  As usual, $C'(r) \approx 1$ while $1\lesssim r \lesssim H_0^{-1}$, after which it approaches its asymptotic value $R$.

\subsection{Limiting case $R= 1$, and the  half-BPS solution}

Finally we discuss the special case where the extra dimension is such that the asymptotic radius of the compact dimension is  $R= 1$. This is
a particularly interesting case, since in our conventions, this corresponds to the natural size
of the vortex in the supersymmetric limit of the theory, namely when $\beta=1$ and $\gamma=1$.

\begin{figure}[t]
\centering \makebox[\textwidth][c]
{\hspace{2.1cm}\includegraphics[width=0.62\textwidth]{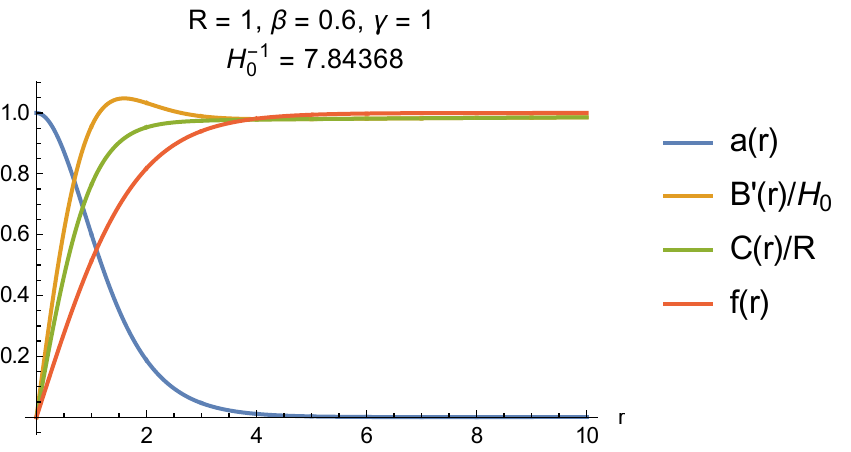}\hspace{0.1cm}\includegraphics[width=0.62\textwidth]{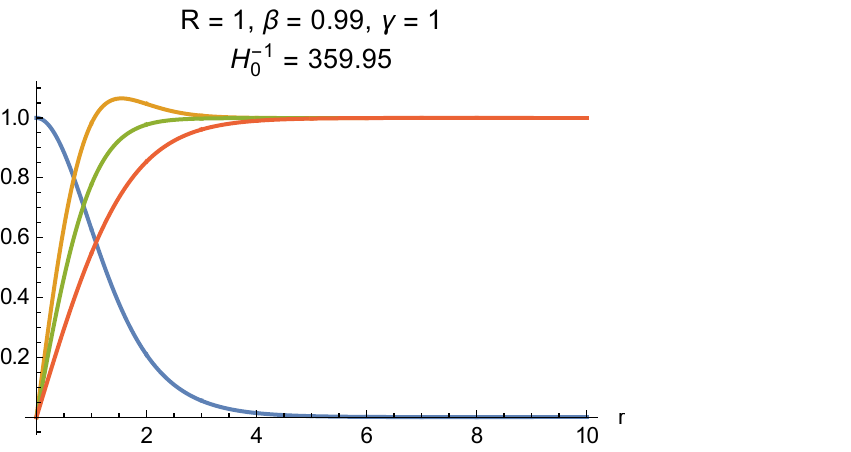}}
\caption{On the left plot we show a bubble with gravitational coupling $\gamma=1$ and KK radius $R= 1$ away from the supersymmetric state, $\beta = 0.6$. Note the important deviation from
the thin wall approximation.  The right plot shows  a bubble with $\gamma=1$ and $R = 1$ near the supersymmetric limit, $\beta = 0.99$.  The radius of the bubble $H_0^{-1}$ becomes very large in this case.  
All indications are that $H_0 \to 0$ as $\beta \to 1$ from below.}
  \label{bubble-C-1}
\end{figure} 
We cannot rely on the thin wall approximation in this case either since there is no real separation
between the size of the vortex and the compact space volume, specially if we take the $\beta<1$ 
case where the vortex core would be even bigger, as follows from the experience in asymptotically  conical spacetimes.
We show an example of these solutions with $\beta=0.6$ in figure~\ref{bubble-C-1}, which proves  the existence of bubble of nothing configurations in this regime.
Nevertheless it is interesting to note that taking the supersymmetric limit one arrives to practically the same conclusions 
as in the previous cases,  as shown in  the right panel of figure~\ref{bubble-C-1}: the bubble becomes flat and infinite, signaling an exponential suppression of the decay rate. \\

Let us now discuss more in detail the limiting solution we obtain for $R=1$, as we approach the supersymmetric case $(\gamma=1, \beta=1)$. In previous sections we have shown that in this limit  the  spacetime metric displays two  different regimes. 
On the one hand, near the string core,  $1\lesssim r\lesssim H_0^{-1}$, the geometry is dominated by the vortex configuration and $C(r) \approx 1$. On the other hand,  at large distances from the core, $r \gg H_0^{-1}$, the metric approaches the asymptotic KK configuration  with radius $C(r) = R$. Our numerical calculations also indicate  that  $H_0$ vanishes in the supersymmetric limit, which implies that the first regime becomes infinite. Interestingly, in the case $R=1$ the near vortex configuration  already  meets the asymptotic boundary condition $C(r) = R =1$, and therefore it would seem that there is no sense in which two regimes are present.  As we shall argue in the following, this is precisely the case at hand. Actually, the resulting configuration 
is a half-BPS vortex solution that is known to exist in our model \cite{Edelstein:1995ba,Dvali:2003zh}.\\

Following the evidence obtained by our previous numerical calculations we will set $H_0$ to zero in order to study the supersymmetric limit. Then, the generalized  ansatz for the metric \eqref{arbitraryR} reduces to  
\begin{equation}
ds^2=B^2(r)\left(- d\tau^2+dz^2\right) +dr^2+C^2(r)d\theta^2 \ .
\label{metricCS}
\end{equation}
Note that  the coordinate $z$ now takes values in the range $(-\infty,\infty)$, implying that the bubble radius is infinite, and therefore the string wrapping it is also infinitely long and flat.  If we  set  $H_0=0$ in the system of equations \eqref{eq:mattereom} and \eqref{eq:Ceom}, it can be shown that they admit a first integral (see for example \cite{Blanco-Pillado:2013axa}), leading to a new system of first order differential equations called the BPS equations,\footnote
{Although we are discussing the supersymmetric limit, we leave here the parameter $\gamma$ explicit for later convenience.}
\be
f' - f a\,  C^{-1}=0 \ , \qquad 
  a' -     C \, (f^2-1)=0 \ ,\qquad 
 C'-1 +   \gamma \left( 1 + a (f^2-1)  \right)=0 \ ,  \label{metricBPSeq}
\ee
while the profile function $B(r)$ can be consistently set to a constant, $B(r)=1$.  
Furthermore, the boundary conditions that impose regularity at the core as well as finite energy (per unit length of the string) reduce to
\be
C(0)=0 \ , \quad  f(0)=0 \  , \qquad  a(0)=1 \ , \qquad f({r \to \infty}) = 1  \ , \qquad    a(r \to \infty) = 0 \ .
\label{bc2}
\ee

\begin{figure}[t]
\centering \hspace{-.5cm}\includegraphics[width=0.8\textwidth]{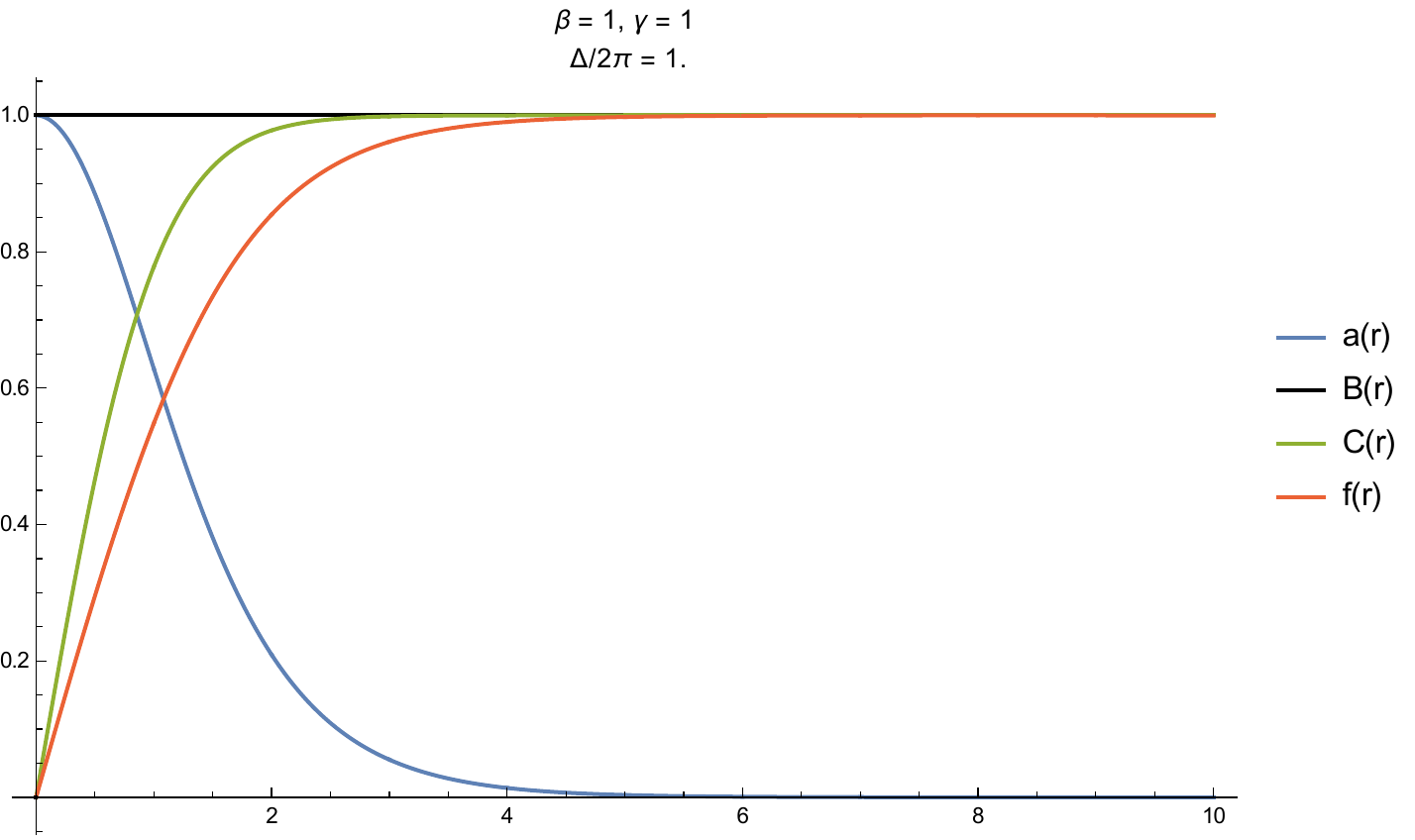}
\caption{Solution in the supersymmetric limit  $\gamma=1$, $\beta = 1$, with asymptotic compactification radius $R=1$. }
  \label{bubble99}
\end{figure}

Field configurations satisfying the BPS  equations and these boundary conditions can be shown to
leave unbroken half of the supersymmetries. More precisely, the unbroken supersymmetries are those generated by a parameter $\epsilon$  of the form
\be
 \epsilon _L(\theta) = \rme^{-\ft12\rmi\theta } \epsilon _{0L},
\label{susyParam}
\ee
where $\epsilon _{0}$  satisfies the projector condition $\gamma^{12} \epsilon_0 = - \rmi \gamma^5 \epsilon_0$ \cite{Edelstein:1995ba,Dvali:2003zh}.   Far away from the core,  $r\to \infty$,  the BPS equation for the metric profile function $C(r)$ becomes 
\be
C' \approx  1-  \gamma  \label{approxMetricBPS},
\ee
therefore,  if we want this spacetime to behave asymptotically as a compactified state
we must require that $C'(r)\to0$ for large $r$, and thus we need the parameters of our theory  to
obey the constraint $\gamma = 1$.  In this case, for $r\to \infty$, the solution approaches the vacuum compactification discussed earlier and 
given  eq.~(\ref{wilsonCylinder}), where supersymmetry is fully restored since the parameter $\gamma$ satisfies the constraint \eqref{susy-condition}. (See \cite{Dvali:2003zh}.) 
Furthermore, the $\theta-$dependence of the supersymmetric 
parameter \eqref{susyParam}   is consistent with the boundary conditions of the fermions, which should 
be anti-periodic, as the spacetime is simply connected.

In figure~\ref{bubble99} we show a numerical solution of the previous system of equations and boundary conditions. Note that the radius of the compactification 
rapidly approaches the asymptotic value $C(r)=R=1$, and therefore this configuration is also a solution of the full set of equations of motion and  boundary 
conditions imposed  in section \ref{sec:bc}, which we used  previously to obtain the bubble configurations. 
 It is interesting to note that the field profiles in this solution are identical to the ones we have found earlier for other values of $R$  near
the supersymmetric limit (see e.g. figure~\ref{bubble-sequence}). This is of course possible because those  solutions
also have a very small $H_0$ in that limit, so the bubble becomes effectively flat, and the outer region $r \gtrsim H_0^{-1}$ is at a very large distance from 
the core. Indeed, as we anticipated at the beginning of this section, in the solution of \eqref{metricBPSeq} and \eqref{bc2} represented in figure~\ref{bubble99}, 
the outer regime is totally absent.\\

The spacetime   in the $(r,\theta)$ directions resembles a cigar type geometry, and 
it is pretty close to that of a cylinder with a spherical cap attached to it on its end \cite{Dvali:2003zh,Blanco-Pillado:2013axa}.
Taking into account this description of the solution, one can interpret the half-BPS cosmic string solution as an
interpolation between two different  vacua. The compactified vacuum  $\mathbb{M}_3\times S^1$ given by eq.~(\ref{wilsonCylinder})
and the pure magnetic spherical compactification $\mathbb{M}_2\times S^2$ at the vortex core, which can also be shown to be a solution
in our model \cite{Dvali:2003zh,Blanco-Pillado:2013axa}.  In this view, the solution is very 
similar to the static supersymmetric domain walls that interpolate between supersymmetric
vacua in supergravity theories \cite{Cvetic:1992st}. The presence
of these half-BPS solutions in those models signals the suppression of a possible decay between such
vacua, which is precisely the behaviour we encounter in our case.  In our model, the decay to the
bubble of nothing is suppressed in this case when $R=1$ by the appearance of a 
half-BPS cosmic string solution that prevents the decay from happening. Other possible initial
supersymmetric configurations are also protected by a similar object, although in these 
cases the solution is not half-BPS due to different asymptotic boundary conditions.

\section{Conclusions}
\label{sec:conc}

Several years ago, Witten showed that compactified higher dimensional theories are susceptible to
decay via the formation of a bubble of nothing. This happens by the spontaneous nucleation of a 
bubble where the extra-dimension pinches off and disappears.  It is generally believed that
supersymmetric compactifications would be stable with respect to this decay channel due to the
necessity of periodic fermions around the extra dimension (circle of compactification). This periodicity
is incompatible with the bubble of nothing cigar geometry, which imposes antiperiodic fermions in the
asymptotic region, the region that approaches the compactification vacuum state. 

On the other hand, this KK vacuum solution is not the only possible compactification on a circle 
that preserves supersymmetry. We have shown in this paper an explicit example of a supersymmetric 
compactification that allows anti-periodic fermions due to the presence of extra matter fields
winding the extra dimension.  It would therefore seem possible for these states to decay via
an instanton similar to the one in the pure Witten bubble.  Here we have investigated this possibility
and concluded that one can indeed find such instantons in those models. The new ingredients in this
vacuum solutions makes it necessary for a vortex to be placed on the instanton geometry in order
to reconcile the asymptotic boundary conditions with the pinching off of the extra dimension. This 
vortex string can be chosen to only mildly deform the geometry when the parameters of the theory are far from
the supersymmetric case, so one can expect these states to be unstable to the formation of the
bubble of nothing. However, in the limit where the compactification is supersymmetric, the solution
is such that the bubble becomes infinite and flat, signaling the suppression of the instability.
This is exactly the same kind of behaviour one encounters in field theory models and
shows that, at least in this case, the suppression of the decay is not due to any topological obstruction or
superselection rule, but rather it has a dynamical origin.  

We have investigated the behaviour of the bubbles within a simple 4D model, where we
can use the thin wall (Nambu-Goto) approximation to estimate the effect of the vortex on the spacetime geometry.
Furthermore, we have done a thorough numerical exploration of this model for different initial
compactification scenarios and parameters, and we have concluded that this effect is realized
in all the cases, even in cases where the thin wall approximation would not be appropriate. This
gives us confidence to speculate that this mechanism is generic. It would be interesting to
investigate the presence of this suppression mechanism in higher dimensional models
of flux compactification in field theory as well as String Theory.

We conjecture that this mechanism stabilizes any supersymmetric compactification
that is not prevented from decay by topological obstructions such as spin structure.

\section*{Acknowledgments}

We would like to thank Tom\'as Ort\'in, Adam Brown, Fernando Quevedo and Miguel Montero. This work was supported in part by IKERBASQUE, 
the Spanish Ministry MINECO (FPA2012-34456) and (FPA2015-64041-C2-1P), the grant from 
the Basque Government (IT-559-10), the Consolider EPI CSD2010-00064, and by the ERC Advanced 
Grant 339169 ``Selfcompletion''.

\appendix

\section{Quantization of the Fayet-Iliopoulos term}

\label{secondappendix}
In the $\cN=1$ locally supersymmetric version of the Abelian-Higgs model, the parameter $\eta^2$, i.e.  the vacuum expectation value of the Higgs field $\phi$, is the so called Fayet-Iliopoulos (FI) term. This parameter also determines  the charge of the gravitino under the local $\U(1)$ gauge transformations and,  as a consequence, it satisfies a quantization condition which takes the form \cite{Distler:2010zg,Seiberg:2010qd}
\be
\eta^2 \kappa^{2} = 2 p, \quad \text{where} \quad p\in \mathbb{Z}.
\label{quantumFIterm}
\ee
For simplicity in the main text we have neglected this condition and treated the FI-term as a continuous parameter, but it is straightforward to  show that our conclusions are not affected when the quantization is taken into account. In particular we will now show that this model admits a supersymmetric compactification of Minkowski space to  $\mathbb{M}_3 \times S_1$ compatible with   anti-periodic boundary conditions for the fermions on the $S^1$.  For this purpose we need to consider a $\cN=1$ locally supersymmetric Abelian-Higgs model  with slightly more general  couplings than the one defined by \eqref{phenomL}. The bosonic sector of the theory we will discuss now is given by
 \begin{eqnarray}
\label{actionapp}
 S_{\textrm{bos}} &=&  \int d^4x \sqrt{- g} \Big[  \frac{1}{2\kappa^2} \, R -  D_{\mu} \bar \phi  \,  D^{\mu} \phi -\frac{1}{4} 
F_{\mu\nu} F^{\mu\nu} -\frac{e^2}{2}\left(\eta^2 - q \phi \bar\phi \right)^2
\label{phenomL2}
\Big], 
\end{eqnarray}
 where the gauge covariant derivative is defined by $D_{\mu} \phi=(\partial_{\mu} - i q e A_{\mu}) \phi$, and the integer $q\in\mathbb{Z}$. Note that, in contrast with the model given by \eqref{phenomL}, the charge of the Higgs is an arbitrary integer multiple $q$ of the gauge coupling $e$. As we shall now see, introducing this new parameter  is essential for the construction of the supersymmetric compactification with anti-periodic fermions. Note also that this model only admits zero-energy vacuum solutions, such as the spontaneous compactification to $\mathbb{M}_3 \times S_1$ that we wish to discuss,  provided  the parameters satisfy
 \be
 \mathrm{sign}(p) =\mathrm{sign}(q),
 \ee
  which we shall assume in the following discussion.  The line element corresponding to the spontaneous compactification  $\mathbb{M}_4\to \mathbb{M}_3 \times S_1$  has the form
\be
 ds^2 = -dt^2 + dz^2 + dr^2 + R^2 d\theta^2,
\label{Appendixds}
\ee
where $R$ is the radius of the compact $S^1$ direction. In addition we will impose anti-periodic boundary conditions for the all the fermions, $\chi$, $\lambda$ and $\psi_\mu$,  along  the compact $S^1$ direction. First we will argue  that this setting is consistent with preserving all the supersymmetries from a  four dimensional point of view. If we restrict ourselves to bosonic configurations  $\psi_\mu = \chi = \lambda =0$, the only non-vanishing  supersymmetry transformations are the ones of the fermions given in  (\ref{gravitinoSUSYtrans}-\ref{gauginoSUSYtrans}). 
Furthermore, we consider the field configuration
\be
\phi = \ft{\eta}{\sqrt{q}}\,  \rme^{\rmi n \theta},  \qquad A_\mu = n /(qe) \, \delta_{\mu \theta},
\label{backgroundApp}
 \ee
which ensures that the supersymmetry transformations of the chiralino $\chi$ and the gaugino $\lambda$ are also zero.  The only remaining supersymmetry transformation is the one of the gravitino \eqref{gravitinoSUSYtrans}, which can also be made  zero provided the supersymmetry  parameter $\epsilon$  satisfies
\be
\cD_\mu \epsilon_L = (\pd_\theta + \frac{p\,  n}{q} ) \epsilon_L =0,  \qquad\Longrightarrow \qquad \epsilon_L(x^\mu) = \rme^{- \rmi \frac{ n\, p}{q} \, \theta} \, \epsilon_L^0 .
\label{gravitinoEqCylinder2}
\ee
From equation \eqref{gravitinoSUSYtrans} it also follows that the supersymmetry parameter must satisfy the same boundary conditions as the gravitino, and thus it must be anti-periodic. Therefore the gravitino equation \eqref{gravitinoEqCylinder2} admits  solutions which are consistent with the boundary conditions provided  the parameters satisfy the relation
 \be
 \frac{|n| p}{q} = \frac{1}{2}.
\label{parameterCritString}
 \ee
As $|p|, |n| \ge1$, it is clear from this relation that  supersymmetry can only be fully preserved  when the charge of the chiral field satisfies $q \ge 2$, which justifies the inclusion of this parameter in \eqref{phenomL2}.
It is worth mentioning that the  relation \eqref{parameterCritString} is precisely the condition which ensures that the locally supersymmetric Abelian-Higgs model  admits a critical cosmic string solutions of winding $n\in \mathbb{Z}$. Such solutions have a deficit angle of $\Delta = 2\pi$, and thus their background geometry  asymptotes to $\mathbb{M}_3 \times S_1$ far away from the centre of the string, where supersymmetry is also fully preserved \cite{Edelstein:1995ba,Dvali:2003zh}.  The deficit angle of these strings solutions is given by (see  \cite{Blanco-Pillado:2013axa})
\be
\Delta = 2 \pi |n| \frac{\eta^2 \kappa^2}{q} = 2\pi,
\ee 
which is equivalent  to the condition \eqref{parameterCritString} when the quantization of the FI term is taken into account. In the rest of the discussion we will assume for simplicity, and without loss of generality, that $n>0$.\\

In \cite{Scherk:1979zr,Scherk:1978ta,Hosotani:1983xw} it was shown that non-periodic boundary conditions for the fermions, or a non-vanishing a vacuum expectation value of the gauge field, could induce masses of the order of the KK scale for the fermions, leading to the  breaking of all the supersymmetries after the dimensional reduction.  However, when both situations occur simultaneously both effects  
may cancel each other,  leading to a supersymmetric dimensionally reduced theory.   We will now argue  that this is precisely the situation we have at hand. Following  \cite{Hosotani:1983xw}, we will discuss the masses  of the Kaluza-Klein modes, and  in particular we show that  the KK spectrum of all fermions contain light modes (with no contribution of the order of the KK scale) when the parameters satisfy the relation \eqref{parameterCritString}. This is a necessary condition for supersymmetry to remain unbroken in the reduced theory, as otherwise we would not have the right spectrum of particles to form the supermultiplets.   The four dimensional fields can be expanded in series of Kaluza-Klein modes as follows:

\bea
\phi &=& \sum_{-\infty}^{\infty}  \phi_{m} \, \rme^{\rmi m \theta}\,, \nonumber\\
  \chi_L &=&\rme^{\frac{\rmi }{2}\theta} \, \sum_{-\infty}^{\infty}   \chi_{m|L}\, \rme^{\rmi m \theta}\,,\nonumber  \\ 
A_\mu &=& \sum_{-\infty}^{\infty}  A_{m|\mu} \, \rme^{\rmi m \theta}\,,\nonumber \\
\lambda_L &=&  \rme^{-\frac{\rmi }{2}\theta} \, \sum_{-\infty}^{\infty}  \lambda_{m|L} \,\rme^{\rmi m \theta}\,,\nonumber\\
 \psi _{\mu L} &=&  \rme^{-\frac{\rmi }{2}\theta} \,  \sum_{-\infty}^{\infty} \psi_{m|\mu L}  \,\rme^{\rmi m \theta}\,, 
\eea
where $m$ is an integer labeling the KK modes, and the fields $\phi_m$, $\chi_{m|L}$, etc...  depend only on the non-compact coordinates $x^a \equiv (t, z,r)$.  For  $A_\mu$ to be real we  also need $A_{m|\mu} = (A_{-m|\mu})^*$. Note that this  ansatz for the KK expansion corresponds to a \emph{generalised dimensional reduction} \cite{Scherk:1979zr,Scherk:1978ta}, which ensures that the fermions satisfy  anti-periodic boundary conditions, e.g.  $\chi_{L}(x^a, \theta + 2 \pi) = - \chi_{L}(x^a, \theta)$. The KK contribution to the masses is determined by the kinetic terms of the fields, and more specifically, by the form of the covariant derivatives along the compact $\theta$ coordinate of the $S ^1$.  Taking into account the quantization of the FI-term the $\U(1)$ gauge transformations  read
\bea
&&\delta_g{\phi} = \rmi q e \phi \, \alpha\,,\nonumber \\
 &&  \delta_g \chi_L = \rmi \left(q+ p \right) e\, \chi_L \,  \alpha\,,\nonumber \\ 
&&\delta_g \psi _{\mu L} = - \rmi   p e   \psi_{\mu L} \, \alpha\,,\nonumber \\
&&\delta_g \lambda_L =-   \rmi p e\lambda_L \, \alpha\,,
\eea
where $\alpha$ is the $\U(1)$ gauge parameter, and thus the covariant derivatives along the $\theta$ direction have the form
\bea
&&D_\theta \phi =(\pd_\theta -  \rmi qe A_\theta) \phi\,,\nonumber \\
 && D_\theta \chi_L =\left( \pd_\theta-  \rmi e (q+ p) A_\theta\right) \chi_L\,,\nonumber  \\ 
&&D_\theta \psi _{\mu L} = (\pd_\theta  +\rmi   ep A_\theta) \psi_{\mu L}\,,\nonumber  \\
&&D_\theta \lambda_L =(\pd_\theta +   \rmi ep A_\theta)   \lambda_L\, .
\eea
Note that, after taking into account the quantization of the FI-term,  all the $\U(1)$ charges are integer multiples of the gauge coupling constant $e$.  Using the expectation value of the gauge boson \eqref{backgroundApp}, we find that the KK modes have a contribution to the mass of the form\footnote{The fields in the chiral and gauge multiplets have additional contributions due to other interactions in the Lagrangian, as  the scalar potential in the case of the chiral field $\phi$. }
\bea
&&M(\phi_m) \sim  |m - n| \, M_{\rm KK}\,,\nonumber \\
&&M(\chi_{m|L}) \sim|m + \ft{1}{2}-  n - \ft{n p}{q} | \, M_{\rm KK}\,,\nonumber  \\ 
&&M( \psi _{m|\mu L}) \sim |m - \ft{1}{2}  +\ft{n p}{q}|\, M_{\rm KK}\,,\nonumber \\
&&M(\lambda_{m|L}) \sim |m - \ft{1}{2} +  \ft{n p}{q}|\, M_{\rm KK}  \, ,
\eea
where the Kaluza-Klein mass scale is set by the radius of the compactification, $M_{\rm KK} = R^{-1}$.  Note that the contributions to the masses arising from the anti-periodicity of the fermions, the $1/2$ terms, are cancelled by the contribution associated to the background gauge field, $np/q$, provided the relation \eqref{parameterCritString} is satisfied. It is now straightforward to check that the $m=n$ KK modes in the chiral multiplet, $\phi_n$ and $\chi_{n|L}$, do not receive contributions to the mass of the order of the KK scale $M_{\rm KK}$. Similarly,  in the gauge and graviton multiplets the $m=0$ modes, $A_{0|\mu}$, $\lambda_{0|L}$ and $\psi _{0|\mu L}$, the Kaluza-Klein contributions to the mass are zero. This completes the consistency check showing that the KK spectrum contains the necessary light  modes to form the supermultiplets of the reduced theory, and thus our results  are  fully compatible with the ones presented in \cite{Scherk:1979zr,Scherk:1978ta,Hosotani:1983xw}.

\section{Numerical solutions}

\label{Numerics}
We obtained numerical solutions to the equations of motion by shooting from the bubble core outward, 
modifying the initial values until the desired asymptotic field values are achieved.  In practice, this must 
be done independently over many adjacent intervals, where intermediate shooting parameters are 
introduced whose values are determined by continuity and smoothness at each junction.  The full
set of shooting parameters is solved for using Newton's method.
This is called 
the multiple-shooting method. Regardless of how shooting is performed, a suitable initial (near core) 
and final (asymptotic) boundary condition must first be obtained.
\subsection{Near core}
Because our equations of motion are singular at the bubble (where $C = 0$), we will first Taylor expand 
all fields about $r=0$, defined as where $C(r)$ vanishes, into generic form.  
A bubble solution without a conical singularity requires fixing the two coefficients 
\begin{align}
C(0)=0, \qquad C'(0)=1,
\end{align} so $C(r) = r + C''(0) r^2/2 + ...$ .  The strongest singularities this introduces into the equations of motion (\ref{eq:mattereom}) and \eqref{eq:Ceom} are
\begin{align}
0&= \frac{\gamma \left(a'(0)^2+4 a(0)^2 f(0)^2\right)}{2 r^2} + {\cal O}(1/r),\\
0&= \frac{a(0)^2f(0)}{r^2} + {\cal O}(1/r),\\
0&= \frac{-\gamma a'(0)^2}{2 r^2} + {\cal O}(1/r),
\end{align}
assuming (as we do throughout this paper) a single winding number for the vortex namely, $n=1$. Recall also that we work in a gauge where  $B(0)=1$. The 
relevant solution to these equations is $a'(0) = 0$, $f(0) = 0$.  At next order, we obtain
\begin{align}
0&=\frac{\left(a(0)^2-1\right) f'(0)}{r} + {\cal O}(r^0),\\
0&=\frac{B'(0)}{r}  + {\cal O}(r^0)
\end{align}
which tells us that (since the sign of $a$ is arbitrary)
\begin{align}
a(0) = 1, \qquad B'(0) = 0.
\end{align}
At next order, we obtain the system of equations
\begin{align}
B''(0)&= \frac{\gamma ( a''(0)^2-\beta)}{4} +  \frac{H_0^2}{2} + {\cal O}(r)\\
C''(0)&=0+ {\cal O}(r)\\
f''(0)&=0+ {\cal O}(r)
\end{align}
Continuing order by order, we are left with three undetermined coefficients, $H_0, f'(0)$, and $a''(0)$.
All three of these should be thought of as shooting parameters, chosen to achieve the three boundary
conditions for $ r \to \infty$,
\begin{align}
a  \to 0,\quad
f  \to 1,\quad
C \to R.\label{eq:bdryconditions}
\end{align} 
Numerically, we can only integrate out to some finite $r = r_{\rm max}$, so we need to match the numerical
solution there onto a suitable asymptotic solution.

\subsection{Asymptotic solution}

We can find an approximate asymptotic solution for $r_{\rm max} \leq r < \infty$ by linearizing 
the matter equations of motion about their vacuum values, yielding
\begin{align}
a''(r) &= 2 a(r) - \left(\frac{2B'(r)}{B(r)} -\frac{C'(r)}{C(r)} \right)a'(r)\\
f''(r) &= 2\beta \left[ f(r)-1\right] - \left( \frac{2B'(r)}{B(r)}+\frac{C'(r)}{C(r)} \right)f'(r).
\end{align}
These can be solved by the WKB method, since at large $r$ the geometrical
coefficients $\left( \frac{2B'(r)}{B(r)} \pm \frac{C'(r)}{C(r)}  \right)$ are small compared to the 
masses $m_a = \sqrt{2}, \,\, m_f = \sqrt{2\beta}$.
By writing $a = \exp(\log a)$ and using the WKB approximation to drop second derivatives 
of $\log a$, we find the second-order equations are well-approximated by the
first-order equations
\begin{align}\label{eq:matterbdrya}
a'(r) &=  - \left[\sqrt{2 + \left( \frac{B'(r)}{B(r)}-\frac{C'(r)}{2C(r)}  \right)^2} + \left( \frac{B'(r)}{B(r)} -\frac{C'(r)}{2C(r)} \right)\right]a(r),\\
f'(r) &=  - \left[\sqrt{2\beta + \left( \frac{B'(r)}{B(r)}+\frac{C'(r)}{2C(r)} \right)^2} + \left( \frac{B'(r)}{B(r)}+\frac{C'(r)}{2C(r)} \right)\right]\left[f(r)-1\right]\label{eq:matterbdryb},
\end{align}
where the signs of the square roots are chosen by the boundary conditions at $r=\infty$.  Of more immediate use, these equations
provide an excellent matter boundary condition for finite $r = r_{\rm max}$, which allows us to use a shooting method to construct
the numerical solutions.  The third boundary condition comes from the (vacuum) constraint equation \eqref{BvsC}, which implies 
\begin{align}\label{eq:matterbdryC}
C(r_{\rm max}) = R\, B'(r_{\rm max}) /H_\delta.
\end{align}
These three relations (\ref{eq:matterbdrya}-\ref{eq:matterbdryC}) are the practical versions of equation (\ref{eq:bdryconditions}).

From equations (\ref{eq:matterbdrya}-\ref{eq:matterbdryb}) it is clear that the matter fields will 
approach their vacuum values exponentially quickly.  Far from the vortex we can trust the
the vacuum Einstein equations, which imply
\begin{align}
r(B)&= r_{\delta} + H_\delta^{-1} \sqrt{B(B - 1)} + H_\delta^{-1}\log\left(\sqrt{B} + \sqrt{B -1}\right)\,,\nonumber\\
C(r)&= R \sqrt{1 -  B(r)^{-1}}\,,
\end{align}
where the parameter
\begin{align}
H_\delta = B'(r_{\rm max})/\sqrt{1-B(r_{\rm max})^{-1}}\,,
\end{align}
is the Hubble parameter of the corresponding pure vacuum bubble of nothing, which would have a delta-function singularity at
\begin{align}
r_{\delta} =r_{\rm max} - H_\delta^{-1} \sqrt{B(B - 1)}|_{r_{\rm max}} - H_\delta^{-1}\log\left(\sqrt{B} + \sqrt{B -1}\right)\Big|_{r_{\rm max}}.
\end{align}

This means that having integrated the solution numerically to a large enough $r_{\rm max}$, we can read
off the parameters of the vacuum solution directly from the numerical values of the functions at $r=r_{\rm max} $.
This method allows us to obtain $r_{\delta}$ and $H_\delta$, and from there we can get the more physical
parameter,  the deficit angle $\Delta_W$. These are related through $R$ by
\begin{align}
H_\delta = \frac{2\pi - \Delta_W}{\pi R},
\end{align}
where $\Delta_W$ is the deficit angle (relative to Witten's bubble of nothing), measured at 
$r = \infty$.  This is the way we obtain the values of $\Delta_W$ that we present in our numerical
solutions and that we compare with the analytic estimates based on the arguments of section \ref{BON-and-STRING}.

\bibliography{supermassive}

\end{document}